\documentclass[prb,final,showpacs,onecolumn,superscriptaddress,aps,5pt]{revtex4-2}
\usepackage[utf8]{inputenc}
\usepackage[normalem]{ulem}
\usepackage{textcomp}
\usepackage{amssymb}
\usepackage{natbib}
\usepackage{amsmath}
\usepackage{amsfonts}
\usepackage{graphicx}
\usepackage{subfigure}
\usepackage{mathrsfs}
\usepackage{dcolumn} 
\usepackage[T1]{fontenc}
\usepackage{bm}
\usepackage{color}
\usepackage{lipsum}
\usepackage{cancel}
\usepackage{mathbbol}
\usepackage{epsfig}
\usepackage{units}
\usepackage{esint}
\usepackage{soul} 
\usepackage{url}
\usepackage{afterpage}
\usepackage{hyperref}
\usepackage{braket}
\newcommand{\me}[1]{\left\langle #1 \right\rangle }

\newcommand{\tr}{\mathrm{tr}}

\begin{document}
\title{Large-$n$ $O(n)$ with long-range interactions: integrability and resonance dynamics} 
	\author{Guido Giachetti}
    \affiliation{Laboratoire de Physique de l'\'Ecole Normale Sup\'erieure, CNRS, ENS $\&$ PSL University, Sorbonne Universit\'e, Universit\'e Paris Cité, 75005 Paris, France}
	\author{Nicol\`o Defenu}
    \affiliation{Institut f\"ur Theoretische Physik, ETH Z\"urich, Wolfgang-Pauli-Str. 27 Z\"urich, Switzerland}
    \affiliation{CNR-INO, Area Science Park, Basovizza, I-34149 Trieste, Italy}
	\begin{abstract}
	\noindent
    We study the large-$n$ dynamics of the long-range quantum $O(n)$ model, focusing on the strong long-range regime $\alpha<d$. The dynamics of the model exhibits non-trivial features on mesoscopic timescales $t\sim\ln N$, due to the activation of parametric resonances of the nearly degenerate quantum modes.  By using recent results establishing the integrability of the large-$n$ limit, we derive the exact resonance conditions and the associated phase diagram, construct the reduced Hamiltonian governing the nonlinear evolution of the resonant sector, and determine the quasiperiodic spectrum of the resulting mesoscopic dynamics. This framework offers a quantitative understanding on how quantum deviation can drive the system away from the mean-field limit. In particular, the presence of multiple resonant modes enhances the logarithmic growth of entanglement and leads to spatially modulated correlations.
 \end{abstract}
	\maketitle 
\noindent

\section{Introduction}
Within the framework of statistical mechanics and condensed matter physics, it is well established that long-range (LR) interactions—i.e., slowly decaying power-law couplings—can give rise to a wealth of novel phenomena in both equilibrium and non-equilibrium regimes, for classical~\cite{campa2014physics} as well as quantum~\cite{defenu2021longrange,defenu2024out} systems. This topic has recently attracted renewed attention, largely due to the increasing feasibility of experimental realizations in atomic, molecular, and optical (AMO) platforms~\cite{haffner2008quantum,lahaye2009physics,saffman2010quantum,ritsch2013cold,bernien2017quantum,monroe2021programmable,defenu2021longrange,mivehvar2021cavity}. These advances have unveiled a rich phenomenology, including dynamical phase transitions~\cite{zhang2017observation,baumann2010dicke}, exotic quantum phases~\cite{rovny2018observation,choi2017observation}, and modified universal defect-scaling laws~\cite{safavi2018verification,keesling2019quantum}. In particular, the dynamical behavior of LR systems has sparked intense interest in recent years, owing to their intrinsic tendency to avoid equilibration~\cite{kastner2011diverging,schutz2014prethermalization,defenu2021metastability,defenu2024out}. This lack of thermalization plays a crucial role in stabilizing non-equilibrium many-body phases, such as discrete and higher-order time crystals~\cite{pizzi2021higher,collura2022discrete,giachetti2022high}.

Within this context, the $d$-dimensional $O(n)$ model—long regarded as a cornerstone of out-of-equilibrium field theory~\cite{hohenberg1977theory,ZinnJustin,sotiriadi2010quantum,uhlmann2010system,uhlmann2010on,gambassi2011quantum}—provides an ideal framework to investigate the impact of long-range interactions. Its critical behavior depends crucially on the spatial decay of the coupling, $J(r) \sim r^{-\alpha}$, where $r$ denotes the distance between lattice sites. In particular, a threshold value $\sigma^* > 0$ exists such that, for $\alpha > d + \sigma^*$, the finite-temperature criticality falls within the short-range universality class, whereas for $d < \alpha < d + \sigma^*$~\cite{sak1973recursion,angelini2014relations,defenu2015fixed,behan2017scaling}—the so-called weak long-range regime~\cite{Defenu2020review}—the critical exponents become $\alpha$-dependent. The threshold $\sigma^*$ increases from $\sigma^* = 1$ in $d = 1$~\cite{anderson1969exact,Dyson1969,cardy1981one,angelini2014relations,benedetti2025one,pagni2025one} to $\sigma^* = 2$ for $d \geq 4$~\cite{sak1973recursion,joyce1966spherical,angelini2014relations,defenu2015fixed}, exhibiting a nontrivial dependence on $n$ in intermediate dimensions~\cite{defenu2015fixed,giachetti2021berezinskii}. Analogous results hold for quantum critical points at zero temperature, whose universal properties are connected to those of anisotropic long-range models via the quantum-to-classical correspondence~\cite{sachdev1999quantum,defenu2016anisotropic}. More generally, the leading-order scaling behavior in the weak long-range regime can be mapped to that of local systems in higher (fractional) effective dimensions~\cite{angelini2014relations,defenu2015fixed,solfanelli2024universality}.

In the strong long-range regime, $\alpha < d$~\cite{defenu2024out}, the model displays a striking non-equivalence between descriptions based on different statistical mechanics ensembles~\cite{campa2009statistical,kastner2010nonequivalence,campa2014physics}. The question of how ensemble inequivalence manifests in the quantum regime has recently attracted growing attention~\cite{defenu2024ensemble}, following the pioneering investigations of Ref.~\cite{kastner2010nonequivalence,kastner2010nonequivalenceb}.

Similarly, the quantum dynamics of the $O(n)$ model far from equilibrium has been widely studied in the large-$n$ limit~\cite{moshe2003quantum}. In this regime, depending on the initial state of the dynamics, the problem can be solved exactly~\cite{sotiriadi2010quantum,uhlmann2010on,gambassi2011quantum} and is described by the spherical model~\cite{berlin1952spherical,kac1971spherical}. The large-$n$ quantum dynamics displays a rich and complex behavior, already for local interactions, including dynamical phase transitions~\cite{sciolla2013quantum,chiocchetta2015shorttime,chiocchetta2016shorttime}, coarsening processes~\cite{sciolla2013quantum,chiocchetta2015shorttime,chiocchetta2016shorttime} and aging phenomena~\cite{maraga2015aging}.

Ref.~\cite{giachetti2021entanglement} investigated the large-$n$ dynamics in the strong long-range regime, showing that large-$n$ $O(n)$ models can be interpreted as an interplay between a classical mean-field degree of freedom—whose oscillations remain undamped—and several quantum bosonic modes responsible for spreading quantum correlations. Although these correlations are suppressed as the system size $N$ grows, parametric resonances in the semiclassical dynamics can amplify them on a mesoscopic timescale $t_{\rm Ehr} \sim \ln N$, during which genuinely quantum effects such as entanglement production occur. Many of these features are shared by other strong long-range models, including long-lived prethermal states~\cite{defenu2021metastability,kastner2011diverging}, persistent oscillations~\cite{lipkin1965validity,lerose2018chaotic,lerose2019impact,giachetti2025conditions}, and anomalous entanglement spreading~\cite{pappalardi2018scrambling,pappalardi2022eigenstate}. However, the relative simplicity of the large-$n$ model enables a semi-analytical description of these phenomena, providing a paradigmatic framework for understanding strong long-range dynamics.

\textcolor{black}{While Ref.~\cite{giachetti2021entanglement} established this physical picture, its description remained largely phenomenological. In particular, although the Floquet analysis correctly identifies the possible emergence of resonant modes, it provides only a local characterization of the dynamics. It is therefore difficult to determine under which conditions resonances are activated, or to describe their nonlinear evolution once they become macroscopically occupied.}

\textcolor{black}{In this work we show that these questions can be addressed by exploiting a remarkable property of the large-$n$ $O(n)$ model, namely its recently uncovered integrability. This structure provides the analytical framework needed to turn this phenomenological picture into a quantitative theory of mesoscopic resonances and of their impact on correlation spreading and entanglement dynamics. Building upon the formalism introduced in Ref.~\cite{giachetti2025universality}, we develop a quantitative theory of the mesoscopic dynamics in the strong long-range regime. This allows us to determine the onset of parametric resonances beyond the sole Floquet analysis, describe their nonlinear evolution through a reduced integrable Hamiltonian, and revisit the resulting entanglement dynamics within a unified framework. Our approach provides fresh insights into this problem, offering a comprehensive and coherent picture of the underlying physics. }

The paper is organized as follows. In Sec.~\ref{sec:model}, we introduce the model and discuss its dynamics in large-$n$ limit, highlighting the physical mechanism behind the onset of resonances. \textcolor{black}{ In Sec.\,\ref{sec:linear} we exploit the integrals of motion to characterize the onset of resonances, while Sec.\,\ref{sec:beyond} is devoted to deriving a reduced (integrable) Hamiltonian governing the nonlinear dynamics of the resonant modes, which allows us to  compute its quasiperiodic spectrum exactly. Finally, Sec.\,\ref{sec:entanglement} revisits the entanglement dynamics in light of this effective description, expanding the analysis of Ref.\,\cite{giachetti2021entanglement} to extensive intervals, and highlighting the qualitative differences between the single- and multi-resonant regimes.
}

\section{The model} \label{sec:model}
We introduce here the quantum $O(n)$ model and discuss the effect of long-range interactions. Without loss of generality, we can restrict ourselves to the $d=1$ case. We thus consider a quantum chain with $N$ sites (where necessary we assume $N$ to be odd): each site $j$ corresponds to a vector variable of bosonic components $n$ $\boldsymbol{\Phi}_{j}$ and its conjugate momentum $\boldsymbol{\Pi}_{j}$ ($[\Phi^{a}_{j},\Pi^{a'}_{j'}] = \text{i} \ \delta_{r,r'} \delta_{j,j'}$, with $a,a'=1, \dots, n$). The dynamics of the model is given by a long-range quadratic hopping and a local, $O(n)$-symmetric, quartic potential
\begin{equation}\label{eq:Ham1}
    H = \sum_{j}   \frac{1}{2} \boldsymbol{\Pi}^2_j + \frac{1}{2 N_\alpha} \sum_{j \neq j'} \frac{\left( \mathbf{\Phi}_j - \mathbf{\Phi}_{j'} \right)^2}{|j-j'|^{\alpha}} + \sum_j \frac{r}{2} \boldsymbol{\Phi}^2_j + \frac{\lambda}{2n} (\boldsymbol{\Phi}^2_j)^2 \, . 
\end{equation}
Here $r$ represents the bare mass of the theory, $\lambda > 0$ the quartic $O(n)$ coupling, $\alpha > 0$ the decay exponent of the interaction and $N_\alpha = \sum_{j>0} j^{-\alpha}$ is the Kac scaling~\cite{kac1963van}. Let us notice that, while for $\alpha > 1$, $N_\alpha$ just fixes the energy scale of the excitation, for $\alpha < 1$ its presence is fundamental to ensure the extensity of the thermodynamic quantities~\cite{campa2014physics}. \textcolor{black}{
Throughout this work we assume periodic boundary conditions. In the presence of long-range interactions, these are implemented by replacing the lattice distance $|j-j^\prime|$ with the corresponding chord distance
\begin{equation}
    d(j,j^\prime) = \frac{N}{\pi} \sin \left(\frac{\pi |j-j^\prime|}{N} \right) \, . 
\end{equation}
}

The range of the interaction can be tuned between the nearest-neighbors limit ($\alpha = \infty$) and the fully connected limit ($\alpha = 0$). For $\alpha > 3$ the hopping term can be seen as a discretized version of the Laplacian operator, so that Hamiltonian~\eqref{eq:Ham1} is the lattice version of the usual short-range $O(n)$ bosonic field theory. For $\alpha < 3$, the ground state of the model undergoes a transition between a massive ($\me{\boldsymbol{\Phi}} \neq 0$) and a massless ($\me{\boldsymbol{\Phi}} = 0$) phase~\cite{sachdev1999quantum}, where $\me{\cdot}$ denotes the expectation value over a quantum state. 

We will consider the following protocol: for $t<0$, we choose $r^{-} > 0$, and assume the system to be in its massless ground state; then at $t=0$ the bare mass is quenched to its (possibly negative) final value $r$ \textcolor{black}{(see also Sec.\,\ref{sec:quench})}. In particular, here we will focus on the strong-long-range regime $\alpha < 1$. 

The dynamics of the model becomes tractable in the limit $n \rightarrow \infty$, under the condition that the initial state is factorized and homogeneous over both the lattice and the component indices~\cite{cugliandolo2013out}: in this case, for $n \rightarrow \infty$, such a factorization is maintained at any time $t$, so that it is possible to replace $\boldsymbol{\Phi}^2_j \Phi^{a}_j \rightarrow n \me{\Phi^2_a} \Phi^{2}_j$ in the Heisenberg equations of motion (e.o.m.) coming from Eq.~\eqref{eq:Ham1}. 

In the large-$n$ limit the e.o.m. can thus be written explicitly in Fourier space  
\begin{equation} \label{eq:eomPhiPi}
 \dot{\Phi}_\nu =  \Pi_\nu^\dagger, \ \ \dot{\Pi}_\nu = - (r + \lambda \me{\Phi^2} + \omega_\nu^2)  \Phi_\nu^\dagger, 
\end{equation}
as each component evolves identically and independently from the others, the internal field index $a\in\{1,\cdots,n\}$ has been dropped. Here $\nu = - (N-1)/2, \cdots, (N-1)/2$, and 
\begin{equation}
    \Phi_\nu = \frac{1}{\sqrt{N}} \sum_j e^{ 2 \pi \text{i} j \nu/N} \Phi_j \hspace{0.5cm} \Pi_\nu = \frac{1}{\sqrt{N}} \sum_j e^{- 2 \pi \text{i} \nu j /N} \Pi_j
\end{equation}
are the bosonic Fourier modes ($[\Phi_\nu,\Pi_{\nu'}] = \text{i} \ \delta_{\nu' \nu}$); $\omega_\nu^2$ is the lattice dispersion relation 
\begin{equation} \label{eq:disperion}
    \omega_\nu^2 = \frac{1}{2 N_\alpha} \sum_{j >0} \frac{1-\cos ( 2 \pi j \nu/N)}{j^\alpha} \, .
\end{equation}
\textcolor{black}{The choice to label the modes through the integer index $\nu$ instead of the Fourier wavenumber $k = 2 \pi \nu/N$ is more natural in the strong long-range regime $\alpha < 1$, as it will become clear in a moment. Moreover, in the following, we will restrict ourselves to $\nu \geq 0$ by exploiting the symmetry $\Phi_{-\nu} = \Phi_{\nu}^\dagger$, $\Pi_{-\nu} = \Pi_{\nu}^\dagger$, which in turn stems from the fact that the original variables $\Phi_j$, $\Pi_j$ were Hermitian. Correspondingly, whenever Fourier sums are written in this reduced notation, the contribution of the degenerate partners is understood to be included through the appropriate multiplicity factors.} 

Physically speaking, we can interpret the system of Eq.\,\eqref{eq:eomPhiPi} as a collection of $N$ bosons, coupled through a collective, time-dependent, classical degree of freedom
\begin{equation}
m^2 (t) = r + \lambda \me{\Phi^2}  = r + \frac{\lambda}{N} \sum_\nu \me{\Phi^\dagger_\nu \Phi_\nu}   
\end{equation}
which plays the role of the renormalized mass. For constant $m$, the system reduces to a set of independent bosonic modes, whereas for a time-dependent $m(t)$, the occupation number of the modes may change and scattering occurs. Notice that $m^2(t)$ need not be positive at any time, while $m^2(t)\geq0$ in any stationary state. 

The dramatic simplification at the level of the dynamics can be appreciated by noticing that the time-evolution of second order correlators can be cast in terms of a closed set of ODE. In particular, if we parameterize the second order moments as $|\eta_\nu |^2 \equiv \me{\Phi^\dagger_\nu \Phi_\nu}$, $2 \Re (\eta^{*}_\nu p_\nu) \equiv \me{\Pi_\nu \Phi_\nu  + \text{h.c.}}$, $|p_\nu|^2 \equiv \me{\Pi^\dagger_\nu \Pi_\nu}$, the e.o.m. \eqref{eq:eomPhiPi} become 
\begin{equation} \label{eq:eometap}
    \dot{\eta}_\nu = p_\nu, \hspace{0.5cm} \dot{p}_\nu = - \left( r + \omega_\nu^2 + \frac{\lambda}{N} \sum_\nu |\eta_\nu|^2 \right) \eta_\nu \, . 
\end{equation}
Let us notice that the range of the interaction enters the e.o.m. only through the dispersion $\omega_\nu$: in particular, for $0 < \alpha < 1$, the spectrum remains discrete even in the thermodynamic limit~\cite{defenu2021metastability}, as the $\omega_\nu$ accumulate toward $\omega_\infty^2 = 1$ as $\nu \rightarrow \infty$. Indeed, for large $N$ Eq.\,\eqref{eq:disperion} can be rewritten (see \cite{defenu2021metastability}) as 
\begin{equation} \label{eq:dispersionLR}
    \omega_{\nu}^2 = 1 - (1-\alpha)  \pi^{\alpha-1} \int_0^{\pi} dx \frac{\cos(\nu x)}{x^\alpha}
\end{equation}
while $\omega_0 = 0$ as required by the translational invariance of the model. In the $\alpha = 0 $ limit, we have $\omega_\nu = 1- \delta_{\nu,0}$, so that we have an almost perfect degeneracy broken only by the zero mode, in agreement with the perfect permutational symmetry of the model. 

\textcolor{black}{The peculiar form of the dispersion of the excitations $\omega_\nu$ in Eq.\,\eqref{eq:dispersionLR} in the strong-long-range regime leads to some further simplification at large $N$. Let us suppose first that all the modes $\eta_\nu$ are $O(1)$: as a growing fraction of the modes lies arbitrarily close to $\omega_\infty = 1$,
\begin{equation} \label{eq:G}
    \frac{1}{N} \sum_\nu |\eta_\nu|^2 \sim  |\eta_\infty|^2 \, .
\end{equation}
Under this assumption, the $ \nu = \infty$ mode will decouple from the others in the large $N$ limit: by introducing $\bar{\eta} = \eta_{\infty}$, $\bar{p} = p_{\infty}$ one has
\begin{equation} \label{eq:quartic_potential}
    \dot{\bar{\eta}} = \bar{p} \ ,\hspace{1cm}  \dot{\bar{p}} = - \left(  r + 1 + \lambda |\bar{\eta}|^2 \right) \bar{\eta} \, ; 
\end{equation}
while the evolution of the $\nu = O(1)$ modes one will be determined by the $|\bar{\eta}|$ through the equation
\begin{equation} \label{eq:Floquet_etap}
    \dot{\eta}_\nu = p_\nu \ ,\hspace{1cm}  \dot{p}_\nu = - \left(  r + \omega_\nu^2 + \lambda |\bar{\eta}|^2 \right) \eta_\nu \, .
\end{equation}} By parameterizing $\bar{\eta} = \xi e^{i \theta}$, we can thus exploit the conservation of the angular momentum $\ell = \xi^2 \dot{\theta}$ to reduce the problem to a one-dimensional radial motion 
\begin{equation} \label{eq:Ueff}
    \epsilon = \frac{1}{2} \dot{\xi}^2 + \frac{\ell^2}{2 \xi^2} + \frac{1}{2} (r+1) \xi^2 + \frac{\lambda}{4} \xi^4 \equiv \frac{1}{2} \dot{\xi}^2 + U_{\rm eff} (\xi)  .
\end{equation}
As $m^2 = r + \lambda \xi^2$, we have that $m(t)$ is periodic, with a period that will be denoted as $\tau$. The angular variable $\theta$ is also periodic with a different frequency.

\textcolor{black}{So far, we have assumed that the modes $\eta_\nu$ with $\nu = O(1)$ were $O(1)$: indeed, if $|\eta_\nu|^2 = O(N)$ for some $\nu$, the basic assumption \eqref{eq:G} is no longer justified. To control whether this is the case, one has to examine the evolution equation \eqref{eq:Floquet_etap} for the $\eta_\nu$ which takes the form 
\begin{equation} \label{eq:FloquetHill}
    \ddot{\eta}_{\nu} = - \left( r + \omega_{\nu}^2 + \lambda \xi^2 (t) \right) \eta_{\nu} .
\end{equation}
As $|\bar{\eta}(t)|^2$ is periodic with period $\tau$, Eq.\,\eqref{eq:FloquetHill} takes the form of a Floquet--Hill equation. Its basic properties are briefly reviewed in Appendix\,\ref{app:Floquet}, but for the present discussion it is sufficient to recall that two qualitatively different behaviors are possible. In the stable regime the fluctuations remain bounded and simply acquire a second Floquet frequency on top of the driving period $\tau$. In the resonant regime, instead, they grow exponentially, $\eta_\nu(t)\sim e^{\lambda_\nu^F t}$, where $\lambda_\nu^F>0$ is the corresponding Floquet exponent. Each resonant mode thus becomes macroscopic on the mesoscopic timescale $t_\nu \sim (2 \lambda_\nu)^{-1} \ln N$, as $|\eta_\nu (t_\nu)|^2 \sim N$, whereas in the stable regime the present description remains valid up to polynomial times in $N$. The timescale $t_{\rm Ehr} \sim \ln N$ is therefore the scale on which the classical approximation breaks down in presence of resonances.}

\textcolor{black}{The physical interpretation of this picture is particularly transparent. In the thermodynamic limit all Fourier components collapse onto the accumulation point, $\eta_\nu\rightarrow\bar{\eta}$, yielding
\begin{equation}
\me{\Phi_{j} \Phi_{j'}} = \frac{1}{N} \sum_\nu |\eta^2_\nu| e^{2 \pi \text{i} \nu (j-j')/N} \sim |\bar{\eta}^2| \delta_{j'j} \, .  
\end{equation}
Since all connected correlations between different lattice sites vanish in this limit, no quantum correlation can arise between different sites, justifying the interpretation of $\bar{\eta}$ as a classical degree of freedom. The remaining modes $\eta_\nu$, on the other hand, describe spatially modulated quantum fluctuations on top of this mean-field trajectory. Their resonant amplification marks the onset of a genuinely quantum mesoscopic regime, in which a finite number of quantum degrees of freedom acquire a macroscopic occupation and progressively invalidate the purely classical description.}

\textcolor{black}{The emergence of quantum resonances on logarithmic mesoscopic timescales has been recognized as a generic feature of strong long-range interacting systems, both in fully-connected spin models \cite{lerose2018chaotic, pappalardi2018scrambling} and, more recently, in the large-$n$ $O(n)$ model itself \cite{giachetti2021entanglement}. Despite diverging with the system size, these times remain remarkably short even for thermodynamically large systems because of their logarithmic dependence on $N$, making the associated physics experimentally and numerically accessible. However, in generic interacting models it is notoriously difficult to determine under which conditions resonances are activated, which modes become unstable, and how the nonlinear dynamics evolves once the linear Floquet description breaks down. As we shall show in Sec.\,\ref{sec:linear} and Sec.\,\ref{sec:beyond}, the integrability of the large-$n$ $O(n)$ model provides a complete framework to address these questions, allowing one to characterize both the onset of the resonances and the effective dynamics beyond the linear regime.}

\subsection{Quench protocol}
\label{sec:quench}
\textcolor{black}{Before proceeding with our analysis it is convenient to examine closely the dynamics of our classical mode $\bar{\eta}$ in the concrete case of a ground-state quench in the bare mass $r$ ($r^{-} \rightarrow r$) we are interested in.} In particular, as we initialize our system in the system ground-state with bare mass $r^-$, one has
\begin{equation}
\me{\Phi_\nu (0) \Phi_\nu (0)^{\dagger}} = \frac{1}{2} (\omega_\nu ^2 + m^2_{\rm gs})^{-1/2}, \ , \hspace{0.3cm}
\me{\Pi_\nu (0) \Pi_{\nu}(0)^{\dagger}} = \frac{1}{2} (\omega_\nu ^2 + m^2_{\rm gs})^{1/2}, \hspace{0.3cm}
\me{\Phi_\nu (0) \Pi_{\nu}(0)} = \frac{\text{i}}{2} \, 
\end{equation}
$m_{\rm gs}$ being the ground-state mass (yet to be determined), so that, up to immaterial phase factors,
\begin{equation}
\bar{\eta} (0) =  \frac{1}{\sqrt{2}} \left( m^2_{\rm gs} + 1\right)^{-1/4},  \hspace{0.5cm} \bar{p} (0) =  \frac{\text{i}}{\sqrt{2}} \left(m_{\rm gs}^2  + 1 \right)^{1/4}
\end{equation}
and $\ell = 1/2$. Finally, $m_{\rm gs}$ can be fixed imposing $m^2 = r^- + \lambda \xi^2$, which gives the self consistent condition 
\begin{equation} \label{eq:mgs}
    m^2_{\rm gs} = r_- + \frac{\lambda}{2} ( m^2_{\rm gs} + 1)^{-1/2} .   
\end{equation}
In terms of the single degree of freedom $\xi$, $\theta$ these considerations have a transparent physical interpretation, as
\begin{equation}
    \xi^2(0) = \frac{1}{2} (m^2_{\rm gs}+1)^{-1/2} ,\hspace{1cm} \dot{\xi}(0) = 0 \, . 
\end{equation}
corresponds to the minimum of the effective radial potential $U_{\rm eff} (\xi)$ in Eq.\,\eqref{eq:Ueff} for $r = r^-$. 
Let us notice that Eq.\,\eqref{eq:mgs} predicts $m_{\rm gs} \rightarrow 0$ as $r_- \rightarrow r^c_{\rm gs} = - \lambda/2$, signaling the onset of a quantum phase transition, characterized by the condensation of one of the modes $\Phi^\alpha$ which spontaneously breaks the $O(n)$ symmetry of the model. In the following, we will assume $r^- > r^c_{\rm gs}$.

\begin{figure}
    \centering
    \includegraphics[width=0.88\textwidth]{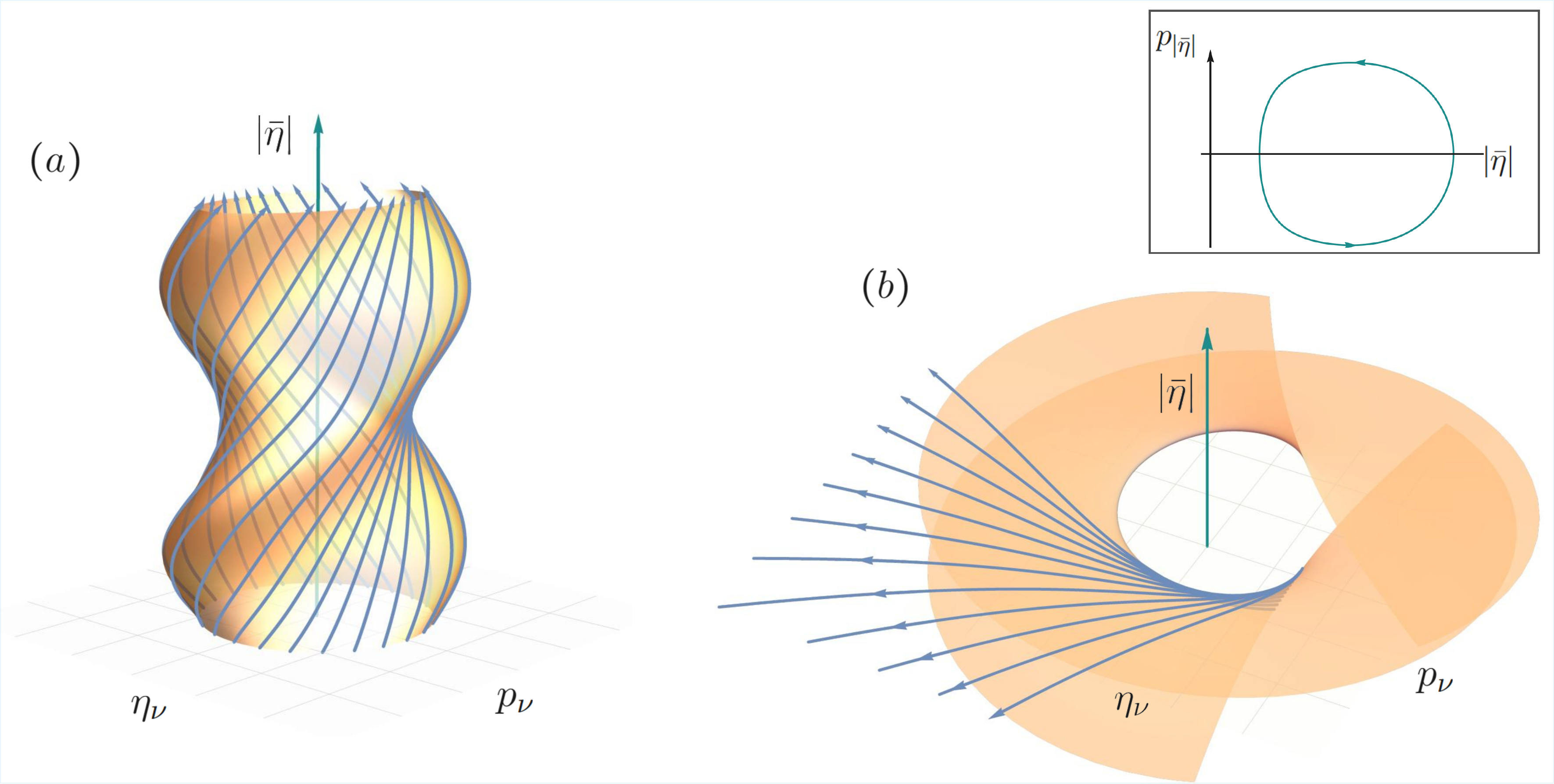} 
    \caption{Structure of the $p_\nu$, $\eta_\nu$, $|\bar{\eta}| = \xi$ section of the phase space in the linear regime $p_\nu,\eta_\nu = O(1)$ in the off-resonant $(a)$ and resonant case $(b)$ respectively. The presence of the integrals of motion constraints the trajectories $(p_\nu, \eta_\nu, |\bar{\eta}|)$ (blue) on the manifold $\epsilon_\nu = \rm const$ (yellow), while the trajectory of the $(p_{|\bar{\eta}|}, |\bar{\eta}|)$ is shown in the inset. While in the off-resonant case the $|\bar{\eta}| = \rm const$ sections of the manifolds are ellipses, preventing the indefinite growth of the $\eta_\nu$, $p_\nu$; if a resonance occurs $\eta_\nu$, $p_\nu$ are no longer bounded as these sections are hyperbolas.} \label{fig:manifolds}
\end{figure}

\section{Integrability: Linear regime}  \label{sec:linear}
\textcolor{black}{The picture emerging from the previous discussion is therefore that the resonant amplification of quantum fluctuations progressively destabilizes the periodic classical motion on the mesoscopic timescale $t\sim\ln N$, marking the breakdown of the classical description. While this mechanism is expected to be rather generic in strong long-range interacting systems, the Floquet approach alone provides only a partial description of the problem. In particular, it does not explain under which conditions resonances are activated, nor does it describe the nonlinear dynamics once the resonant modes become macroscopic and the linear approximation breaks down.
\\\\
These questions can instead be addressed by exploiting a remarkable property of the large-$n$ $O(n)$ model. As recently shown in Ref.~\cite{giachetti2025universality}, its dynamics is in fact integrable, providing a natural analytical framework to investigate both the onset of resonances in the linear regime and their subsequent nonlinear evolution. Indeed, the equation of motion \eqref{eq:eometap} of the $\eta_\nu$, $p_\nu$ can be seen as the Hamilton equations of the classical Hamiltonian}
\begin{equation} \label{eq:classicalH}
\mathcal{H} = \frac{1}{2} \sum_\nu \left(|p^2_{\nu}| + (r + \omega^2_\nu) |\eta^2_{\nu}| \right) + \frac{\lambda}{4N} \left( \sum_{\nu} |\eta^2_\nu| \right)^2.
\end{equation}
In addition to $\mathcal{H}$ itself, the dynamics is constrained by the presence of $(N-1)/2$ trivial integrals of motion, namely $\ell_\nu = \Im(\eta^{*}_\nu p_\nu)$,  $\forall \nu = 1, \dots, (N-1)/2$ linked to the global $U(1)$ symmetry of the Hamiltonian, interpretable as classical angular momenta.  

On the other hand, like its spherical counterpart introduced by C. Neumann~\cite{neumann1859problemate,neumann1856problemate, uhlenbeck2006equivariant}, Hamiltonian~\eqref{eq:classicalH} is known to actually admit a second set of $(N+1)/2$ integrals of motion~\cite{choodnovsky1978completely,wojciechowski1985integrability} in the form of dressed single-mode energies
\begin{equation} \label{eq:iom}
\begin{split} 
    \epsilon_\nu = \frac{1}{2} \left(|p^2_{\nu}| + (r + \omega^2_\nu) |\eta^2_{\nu}| \right) + \frac{\lambda}{4N} |\eta^2_{\nu}| \sum_{\mu} |\eta^2_\mu| +\frac{\lambda}{4N} \sum_{\mu \neq \nu} \frac{|p^2_{\mu} \eta^2_{\nu}| + |p^2_{\nu} \eta^2_\mu| - 2 \Re{(\eta_\nu^{*} p_\nu) } \Re{(\eta_\mu^* p_\mu} )}{\omega_{\mu}^2- \omega_\nu^2} ,
\end{split}
\end{equation}
for $\nu =0, \dots, (N-1)/2$, which are analogous to the so-called Uhlenbeck integrals for the Neumann model~\cite{uhlenbeck2006equivariant}. Recently this formalism has been applied to the study of the classical dynamics of the disordered model~\cite{cugliandolo2018quenched,barbier2019pre, barbier2020non,barbier2022generalised}. Notice that $\mathcal{H} = \sum_{\nu} \epsilon_{\nu} = N \epsilon$. The conserved quantities $\ell_\nu$, $\epsilon_\nu$ provide a complete set of integrals of motion, proving the integrability of our model.
\\\\
\textcolor{black}{We will now show how, in our context, integrability provides a straightforward way to characterize the onset of resonances. To this end, we first focus on the linear regime, namely the stage in which all quantum fluctuations remain microscopic, $\eta_\nu,p_\nu=O(1)$, so that Eq.~\eqref{eq:FloquetHill} provides an accurate description of the dynamics. This assumption is always satisfied at sufficiently short times, irrespective of the presence of resonances, since resonant modes become macroscopic only after the mesoscopic timescale $t\sim\ln N$. From the perspective of the integrals of motion, it means that all the sums in Eq.\,\eqref{eq:iom} can be approximated by their leading term coming from $\mu = \infty$, namely 
\begin{equation}
   \frac{1}{N} \sum_{\mu \neq \nu} \frac{|\eta_\mu^2|}{\omega_\mu^2-\omega_\nu^2} \sim \frac{|\bar{\eta}^2|}{1-\omega_\nu^2} \, , 
\end{equation}
and similarly for the other terms.} It follows that the $\epsilon_\nu$ can be rewritten as
\begin{equation} \label{eq:iomquadratic}
    \epsilon_\nu = \frac{1}{2} \begin{pmatrix} p_\nu^*, \eta_\nu^* \end{pmatrix} \begin{pmatrix}
        U_\nu (\bar{\eta}) & - Q_\nu (\bar{\eta},\bar{p})\\ -Q_\nu (\bar{\eta},\bar{p}) & T_\nu (\bar{\eta},\bar{p})
    \end{pmatrix} \begin{pmatrix} p_\nu \\ \eta_\nu \end{pmatrix} 
\end{equation}
where
\begin{equation} 
     U_\nu (\bar{\eta}) = 1 + \frac{\lambda}{2} \frac{|\bar{\eta}|^2}{1 - \omega^2_\nu }, \hspace{0.3cm}
     T_\nu (\bar{\eta}, \bar{p}) = r + \omega^2_\nu + \frac{\lambda}{2}  \frac{|\bar{p}|^2}{1 - \omega^2_\nu } + \frac{\lambda}{2} |\bar{\eta}^2|, \hspace{0.3cm} Q_\nu (\bar{\eta},\bar{p}) =  \frac{\lambda}{2} \frac{\Re \left( \bar{\eta}^{*} \bar{p} \right)}{1 - \omega^2_\nu } \, . 
\end{equation}
Let us notice that the classical mode $\bar{\eta}$ is associated as well as to a pair of conserved quantities, namely the angular momentum $\ell$ and the energy $\epsilon$ \eqref{eq:Ueff} of the central motion \eqref{eq:quartic_potential}. The determinant $\Delta_\nu$ of the quadratic form in Eq.~\eqref{eq:iomquadratic} is also constant and only depends on $\ell$ and $\epsilon$: 
\begin{equation} \label{eq:Deltanu}
 \Delta_\nu = U_\nu T_\nu - Q_\nu^2 =  r + \omega_\nu^2 + \frac{\lambda \epsilon}{1-\omega^2_\nu} + \frac{\lambda^2}{4} \frac{\ell^2}{(1 - \omega^2_\nu)^2} \, .
\end{equation}
Geometrically speaking, we can interpret \eqref{eq:iomquadratic} by imagining each mode is confined on a manifold that changes periodically in time with $\bar{\eta}$.  \textcolor{black}{The sections of such manifold for $t= \rm const$ are given by conic curves, see Fig.~\ref{fig:manifolds}, as Eq.\,\eqref{eq:iomquadratic} for fixed $\bar{\eta}$, $\bar{p}$ defines an ellipse or a hyperbola in the $|\eta_\nu|$, $|p_\nu|$ plane, depending on the sign of $\Delta_\nu$, as it follows from the diagonalization of the associated quadratic form} \footnote{ \textcolor{black}{More precisely, let us consider the curve defined on $\mathbf{x} = (x_1,x_2) \in \mathbb{C}$ through a quadratic form $\textbf{x}^\dagger A \textbf{x} = \rm const \neq 0$, where $A$ is a real symmetric matrix. The curve be rewritten by introducing a rotated set of coordinates $(X_1,X_2)$ in which $A$ is diagonal, as $\mu_1 |X_1^2| + \mu_2 |X_2^2| = \rm const$, where $\mu_2 \geq \mu_1$ are the eigenvalues of $A$. If $\det A = \mu_1 \mu_2 > 0$, this implies $\mu_1, \mu_2 > 0$ (up to a global sign shift) and the curve is thus an ellipse on the $(|X_1|,|X_2|)$ plane, or an empty set; if $\det A = \mu_1 \mu_2 < 0$ instead, $\mu_1 < 0$ and the curve is a hyperbola.}} \cite{axler2015linear,leon2015linear}. The stability of this motion is thus decided by the sign of the determinant $\Delta_\nu$: positive $\Delta_\nu$ corresponds to an elliptic stable manifold, while in the hyperbolic case $\Delta_\nu$ resonances will occur.

In order to make this picture more quantitative, we introduce the new coordinates 
\begin{equation}
    \eta_\nu = \sqrt{U_\nu (\bar{\eta})} \ q_\nu \, , 
\end{equation}
in terms of which the constants of motion take the diagonal form
\begin{equation} \label{eq:epsilonqU}
    \epsilon_\nu = \frac{1}{2} \left( U^2_\nu (\bar{\eta}) \ |\dot{q}_\nu^2| + \Delta_\nu |q_\nu^2| \right) \, \, . 
\end{equation}
\begin{figure} 
    \centering
    \includegraphics[width=0.49\textwidth]{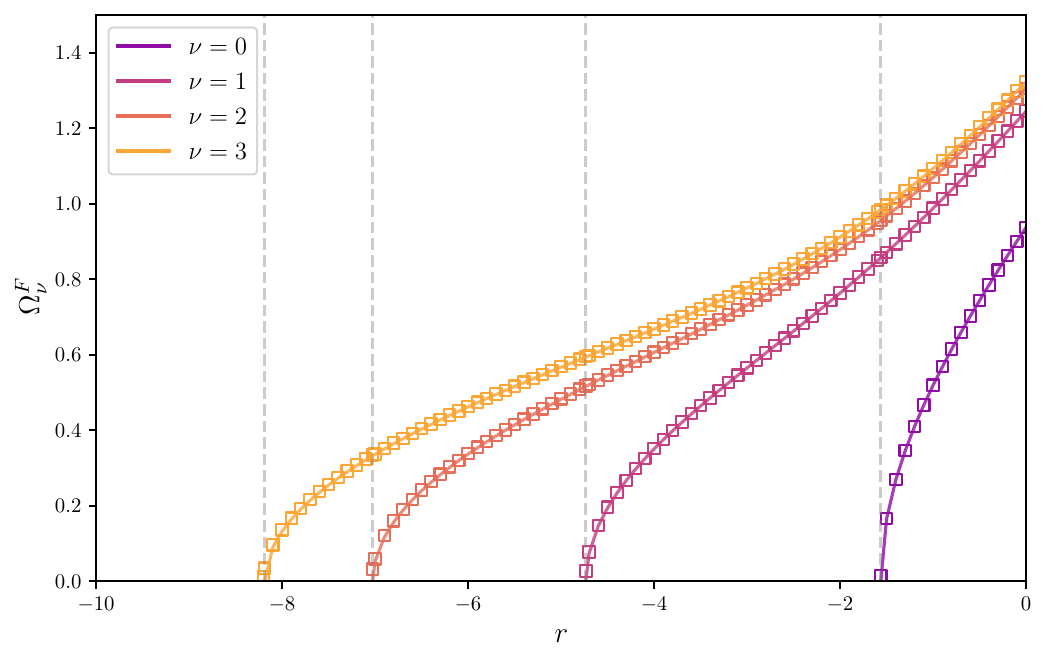} \
    \includegraphics[width=0.49\textwidth]{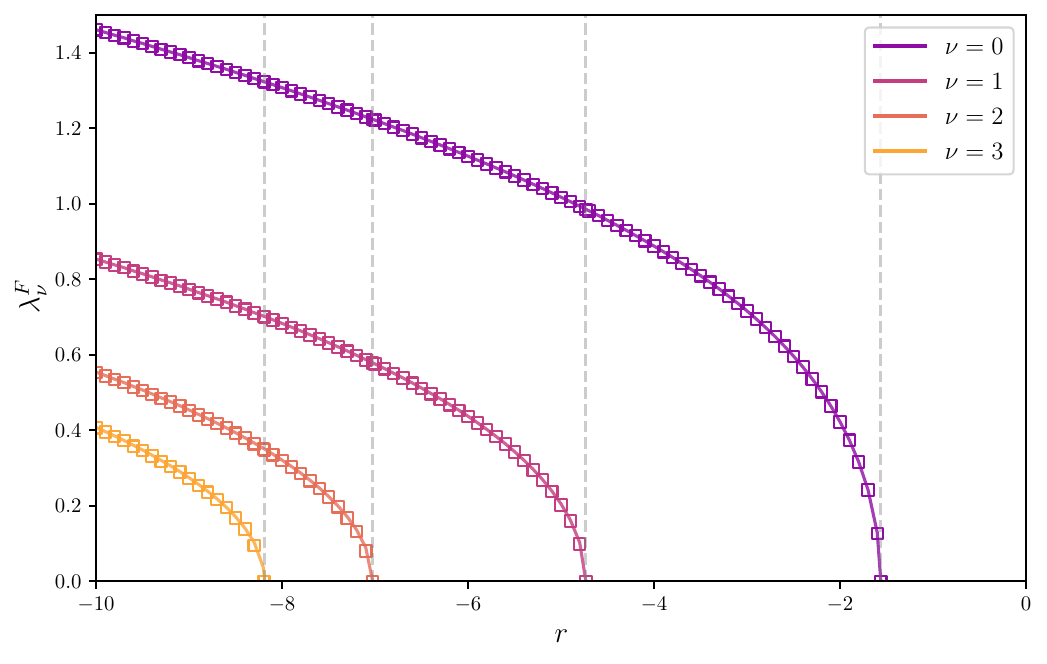}
    \caption{Numerical estimate (diamonds) vs the theoretical estimate (Eq.\,\eqref{eq:FloEig}) of the Floquet eigenvalues $\lambda^F_{\nu}$ corresponding to the Fourier modes $\nu = 0, \dots, 3$ in the  off-resonant (\emph{left}) and resonant (\emph{right}) regimes respectively, for a ground state quench ($m_{\rm gs} = 6$) as a function of the final bare mass $r$. \textcolor{black}{The numerical estimate for the Floquet eigenvalues are obtained by numerically integrating the two fundamental solution of Eq.\,\eqref{eq:FloquetHill} over a period $\tau$ of the forcing and diagonalizing the corresponding monodromy matrix (see Appendix\,\ref{app:Floquet}). In turn $\tau$ is obtained by numerically integrating Eq.\,\eqref{eq:quartic_potential}.} The vertical dotted lines represent the theoretical estimate of the boundaries $r^*_\nu$ of the resonant regions (see Eq.\,\eqref{eq:r*nu}). For the off-resonant regime ($r > r^*_\nu$, \emph{left}) the Floquet frequencies $\Omega^F_\nu = - \text{i} \ \lambda^F_{\nu}$ are shown. } \label{fig:FloEig}
\end{figure}
The $\eta_\nu$ can be written explicitly in the linear regime as a function of the mean-field degree of freedom $\bar{\eta}$ as  
\begin{equation} \label{eq:etalinear}
    \eta_{\nu} (t) = \sqrt{\frac{ U_{\nu}(t)}{ 2\Delta_{\nu}}} \left( A_{\nu}^{+} \ e^{ \sqrt{-\Delta_{\nu}} \int^t_0 dt' \ U^{-1}_{\nu}(t')} - A_{\nu}^{-} \ e^{- \sqrt{-\Delta_{\nu}} \int^t_0 dt' \ U^{-1}_{\nu}(t')}  \right) \, , 
\end{equation}
where $A^{+}_{\nu}$ and $A^{-}_{\nu}$ are constants \textcolor{black}{(notice that, as expected, we have an exponential growth for $\Delta_\nu < 0$ and a quasi-periodic bounded evolution for $\Delta_\nu > 0$)}. As expected, the solution \eqref{eq:etalinear} has the form of a Bloch-Floquet wave, where
\begin{equation} \label{eq:FloEig}
    \lambda^F_\nu = \sqrt{-\Delta_{\nu}} \int^\tau_0 \frac{dt}{\tau} \ U^{-1}_{\nu}(t)
\end{equation}
are the eigenvalues of the Floquet-Hill problem, which become purely imaginary for the off-resonant case. The theoretical estimate of Eq.~\eqref{eq:FloEig} is compared with its numerical counterpart in Fig.~\ref{fig:FloEig}, finding excellent agreement. In particular, in the off-resonant case $\Delta_\mu >0$ ($\lambda^F_\nu = \text{i} \ \Omega_\nu^F$ imaginary) we have that the constants are linked to the integral of motion as
\begin{equation}
   \frac{1}{2} \left( |A^{+}_{\nu}|^2 + |A^{-}_{\nu}|^2 \right) = \epsilon_{\nu} \hspace{1cm} \frac{1}{2\sqrt{\Delta_{\nu}}} \left( |A^{+}_{\nu}|^2 -|A^{-}_{\nu}|^2 \right) = \ell_{\nu} 
\end{equation}
while, if $\Delta_{\nu} < 0$ and the mode is resonant ($\lambda^F_\nu$ real), one has 
\begin{equation}
    \Re \left( A^{+*}_{\nu} A^{-}_{\nu} \right) = \epsilon_{\nu} , \hspace{1cm} \frac{1}{\sqrt{|\Delta_{\nu}|}} \Im \left( A^{+*}_{\nu} A^{-}_{\nu} \right) = \ell_{\nu} \, .
\end{equation}
In this latter case the resonance will thus activate on a timescale $t_{\nu} = (2 \lambda_{\nu})^{-1} \ln N$, so that the period after which the system leaves the linear regime is 
\begin{equation}
    t_{\rm lin} =  \min_\nu (2 \lambda_{\nu})^{-1} \ln N \, .
\end{equation}
Within the context of long-range physics, this can be also interpreted as the Ehrenfest timescale $t_{\rm Ehr}$ for which the semiclassical-mean field approximation is valid\,\cite{lerose2020origin,Defenu2020review}: in the non-resonant regime one has instead that $t_{\rm Ehr}$ is polynomial in $N$, as this is the time-scale in which the incoherent dynamics of non-resonant quantum fluctuations can affect the dynamics of macroscopic observables. \textcolor{black}{It is worth emphasizing that the presence of Floquet resonances does not imply an instability of the full dynamics. Rather, integrability ensures that the motion remains bounded and quasi-periodic, so that the exponential growth predicted by the Floquet analysis only signals the breakdown of the linear approximation around the classical trajectory. The fate of the resonant modes will be discussed in Sec.\,\ref{sec:beyond}.}

\subsection{Stability analysis for the quench protocol}
It is instructive to take a closer look at the stability condition $\Delta_\nu > 0$, namely  
\begin{equation} \label{eq:det>0}
   \Delta_{\nu} = r + \omega_{\nu}^2 + \frac{\lambda \epsilon}{1 - \omega_{\nu}^2} + \frac{\lambda^2 \ell^2}{4(1-\omega_{\nu}^2)^2} > 0 
\end{equation}
\begin{figure} 
    \centering
    \includegraphics[width=0.49\textwidth]{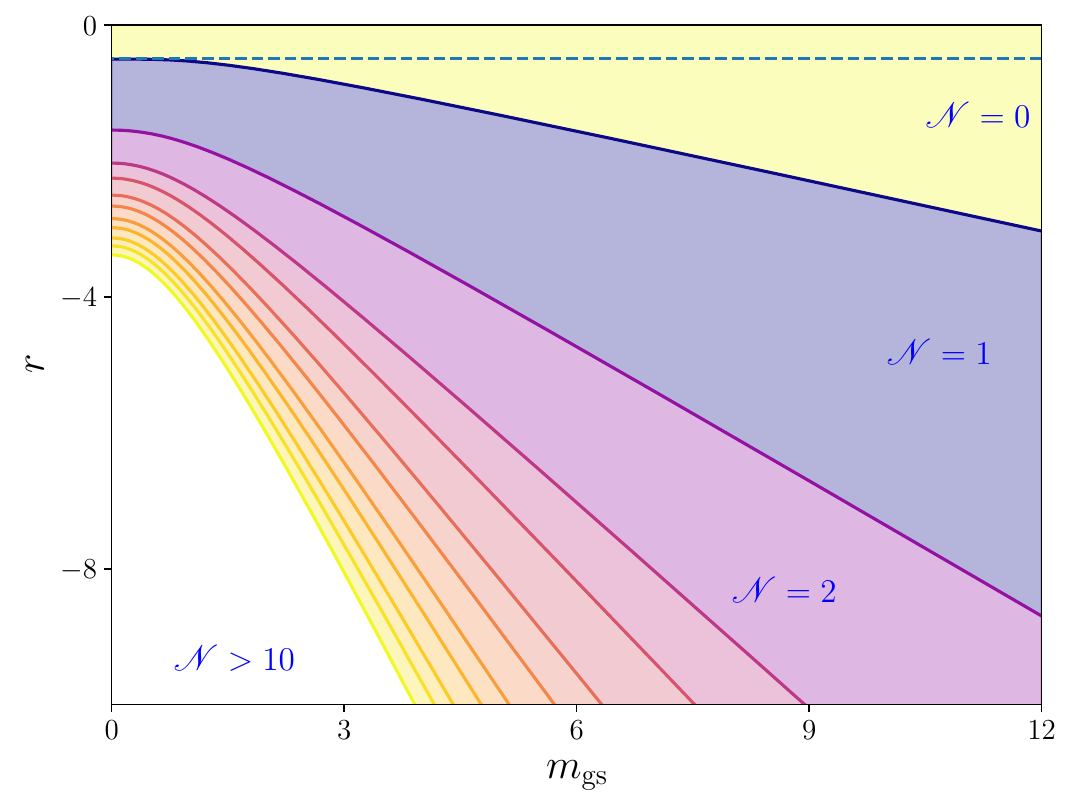} \
    \includegraphics[width=0.49\textwidth]{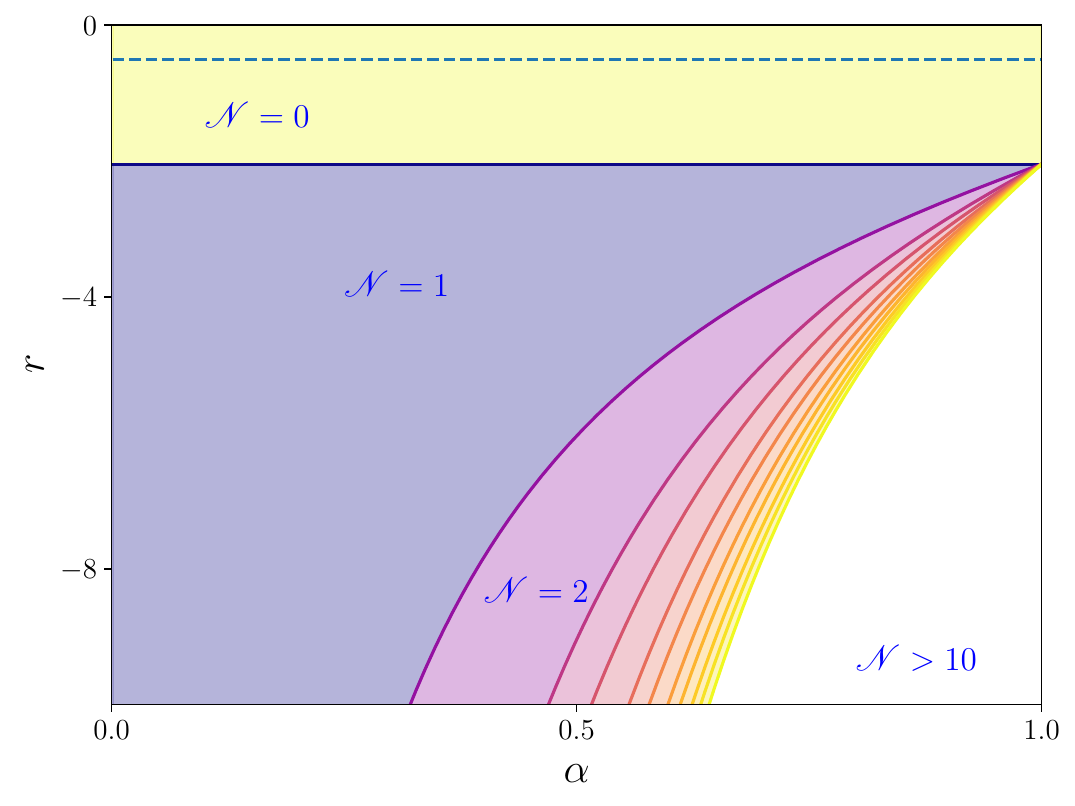}
    \caption{ Phase diagram of the number of resonances $\mathscr{N}$ \textcolor{black}{(obtained by evaluating $\text{sign} \ \Delta_\nu$ for $\nu = 0, \cdots, 9$)}, for a ground state quench $ r_- \rightarrow r$ from the massive phase as a function \textit{(left)} of the ground-state mass $m_{\rm gs}$ and of the post-quench bare mass $r$ for $\lambda = 1$, $\alpha = 0.5$; \textit{(right)} of the range $\alpha$ of the interaction and of the post-quench bare mass $r$ for $\lambda = 1$, $m_{\rm gs} = 8$. The unbroken lines separating the boundaries correspond to the values of $r_\nu^{*}$ in Eq.\,\eqref{eq:r*nu}. We see that the boundary $r^{*}_0$ of the off-resonant region is always below the thermal critical value $r_c^{\rm th}$ (dotted blue line), with $r^{*} \rightarrow r_c^{\rm th}$ as $m_{\rm gs} \rightarrow 0$. Further decreasing $r$ more and more resonances are triggered, with the system becoming less and less stable. \textcolor{black}{In particular, as $\alpha \rightarrow 1$, as the spectrum $\omega_\nu^2$ becomes continuous, our picture breaks down, signaling the onset of the weak-long-range regime in which the mean-field picture is completely lost.} } \label{fig:PhaseDiag}
\end{figure}
for the case of the ground-state quench protocol. First, we notice that $\Delta_\nu$ can be conveniently rewritten by introducing the turning point $\xi_{-}$ of the classical dynamics of $\xi$, i.e. $U_{\rm eff} (\xi_{-}) = \epsilon$, see Eq.\,\eqref{eq:Ueff}. We have 
\begin{equation}
    \Delta_{\nu}  =  \left(1 + \frac{\lambda \xi_{-}^2}{2(1-\omega_{\nu}^2)} \right) \left( r - r^*_0 +  \omega_{\nu}^2 + \frac{\lambda \ell^2 \omega_{\nu}^2}{2\xi_{-}^2(1-\omega_{\nu}^2)} \right) 
\end{equation}
where $r_0^* = - \lambda \epsilon - \frac{\lambda^2}{4} \ell^2$. Thus, the $\nu=0$ mode becomes resonant for $r < r^*_0$, while further decreasing $r$ we encounter a series of critical values $r_{\nu}^*$ at which the mode $\nu > 0$ activates, until the number of resonances becomes extensive as $r \rightarrow - \infty$. Integrability prevents high-energy modes from resonating before the low-energy ones, making the instability of the classical dynamics an intrinsically infrared phenomenon.  

In particular, during a ground state quench $\ell = 1/2$ and $\xi_{-} = \xi(0)$, the zero mode activates at 
\begin{equation}
    r^*_0 = - \frac{\lambda}{4} \left( (1+m^2_{\rm gs})^{1/2} + (1+m^2_{\rm gs})^{-1/2} \right) 
\end{equation}
while for $\nu > 0$
\begin{equation} \label{eq:r*nu}
    r^{*}_{\nu} = r^{*}_{0} - \omega_\nu^2 \left( 1 + \frac{\lambda}{4(1-\omega_{\nu}^2)} \sqrt{1 + m_{\rm gs}^2} \right)  \, . 
\end{equation}
Notice that $ r_0^{*} \leq r^c_{\rm gs} = - \frac{\lambda}{2} $, so that the quench leads to a resonant behavior only if $m^2(t=0^{+}) < 0$. The phase diagram of the number of resonances $\mathscr{N}$ of the model is shown in Fig.\,\ref{fig:PhaseDiag}, both for fixed $\alpha$ and as a function of $\alpha$. 

If no resonance occurs ($\mathscr{N} = 0)$, the correlations do not spread. If only the $\nu = 0$ mode is resonating ($\mathscr{N}=1$), instead, long-range order appears, as a spatially constant term is generated in the correlations. Let us notice that $\mathscr{N}=0$, $\mathscr{N}=1$ phases are the only compatible with an effective permutational symmetry of the system, which is not exact at the level of the Hamiltonian for $\alpha \neq 0$, but it is restored at the dynamical level. These are the only surviving phases in the $\alpha \rightarrow 0$ limit (see Fig.~\ref{fig:PhaseDiag}), where the permutational symmetry becomes exact.

\section{Integrability: Beyond the linear regime}  \label{sec:beyond}

\textcolor{black}{While the previous section established the conditions under which resonances are activated, we now investigate their fate beyond the mesoscopic timescale $t_{\rm Ehr}\sim\ln N$, where the linear approximation breaks down. At this stage the resonant modes have reached macroscopic occupations and can no longer be treated as small fluctuations around the mean-field trajectory. However, since all non-resonant modes remain microscopic, the dynamics is still effectively confined to the finite-dimensional manifold spanned by the classical mode and the resonant modes. We denote by $\mathcal N$ the number of resonant modes, to distinguish it from the system size $N$.} 

Let us thus assume we are in a phase with $\mathscr{N}$ resonances; it is convenient to redefine $p_{\nu} = \sqrt{N} \ \bar{p}_{\nu}$, $\eta_{\nu} = \sqrt{N} \ \bar{\eta}_{\nu}$, with $\nu=0, \cdots, \mathscr{N}-1$. In this notation, the classical mode $\bar{\eta}$ becomes just one of the activated modes, corresponding naturally to $\omega_\infty = 1$, so that it is also convenient to denote $\bar{\eta}_\mathscr{N} \equiv \bar{\eta}$,$\bar{p}_\mathscr{N} \equiv \bar{p}$, $\omega_\mathscr{N} = 1$. \textcolor{black}{As the resonant modes are going to dominate the dynamics,} the Hamiltonian $\mathcal{H}$ in Eq.\,\eqref{eq:classicalH} can be reduced to the $\mathscr{N}+1$ resonant degrees of freedom $\bar{\eta}_\nu$, $\bar{p}_\nu$, obtaining
\begin{equation} \label{eq:mathcalHRes}
    \mathcal{H}_\mathscr{N} =  \frac{1}{2} \sum_{\nu=0}^{\mathscr{N}} \left(|\bar{p}_\nu^2| +  (r+ \omega_\nu^2) |\bar{\eta}_\nu^2| \right) + \frac{\lambda}{4} \left( \sum_{\nu=0}^{\mathscr{N}} |\bar{\eta}_\nu^2|\right)^2   \, 
\end{equation}
(notice that here the parameter $\lambda$ appears without the rescaling with $N$). On the other hand, since $\bar{p}_{\nu} (0), \bar{\eta}_{\nu} (0) = O(N^{-1/2})$, all the integral of motion $\epsilon_{\nu}$, $\ell_{\nu}$ with $\nu=0, \dots, \mathscr{N}-1$ are now $O(N^{-1})$, while  $\epsilon_\infty = \epsilon$, $\ell_{\infty} = \ell$ are $O(1)$. In particular, one has $\ell_\nu / N = \Im(\bar{\eta}^*_\nu \bar{p}_\nu)$ and
\begin{equation} 
\begin{split}
    \frac{\epsilon_\nu}{N} = \frac{1}{2} \left(|\bar{p}^2_{\nu}| + (r + \omega^2_\nu) |\bar{\eta}^2_\nu| \right) + \frac{\lambda}{4} |\bar{\eta}^2_{\nu}| \sum_{\mathbf{\mu}=0}^{\mathscr{N}} |\bar{\eta}^2_\mu| +\frac{\lambda}{4} 
     \sum_{\mathbf{\mu \neq \nu}}^{\mathscr{N}} \frac{|\bar{p}^2_\mu \bar{\eta}^2_{\nu}| + |\bar{p}^2_{\nu} \bar{\eta}^2_{\mu}| - 2 \Re{(\bar{\eta}_\mu^{*} \bar{p}_\mu) } \Re{(\bar{\eta}_\nu^{*} \bar{p}_\nu } )}{\omega_\mu^2- \omega_\nu^2} .
\end{split}
\end{equation}
\begin{figure}
    \centering
    \includegraphics[width=0.69\textwidth]{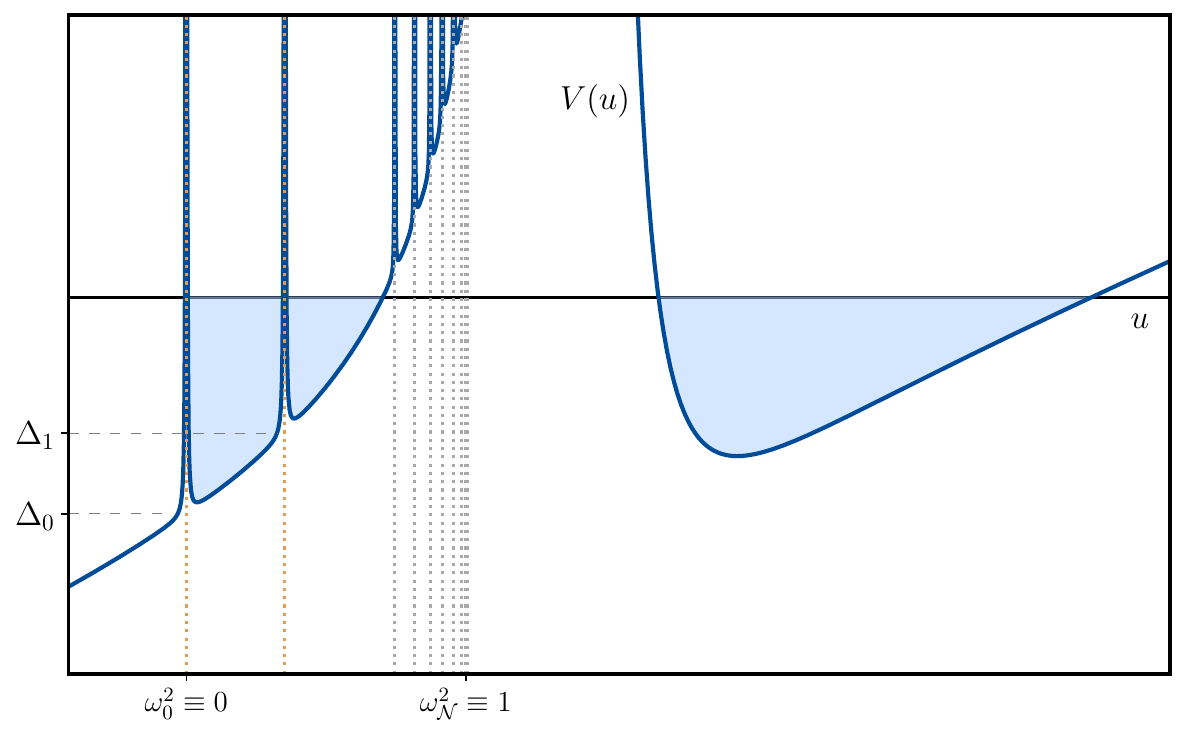} 
    \caption{Sketch of the dynamics of the Jacobi quasiparticles of the long-range $O(n)$ model in the case $\mathscr{N} = 2$. Here the dotted lines represent the dispersion $\omega_\nu^2$ of the modes, the activated ones, $\nu = 0,1$, with $\Delta_\nu < 0$ being in yellow. Each resonant mode corresponds to a Jacobi quasiparticle, which oscillates within the corresponding accessible region, while the last quasiparticle ($u_\mathscr{N} > \omega_\mathscr{N}^2 \equiv 1$) corresponds instead to the classical degree of freedom. The poles corresponding to the off-resonant modes, absent from Eq.\,\eqref{eq:V(u)}, are here shown to clarify their irrelevance in the thermodynamic limit.} \label{fig:Potential}
\end{figure}
\textcolor{black}{Since the reduced Hamiltonian remains integrable, its dynamics is necessarily quasi-periodic and every observable can be expressed as the superposition of $\mathcal N+1$ fundamental frequencies. One of them corresponds to the original mean-field oscillation and remains $O(1)$, while the remaining $\mathcal N$ are associated with the resonant degrees of freedom and are expected to scale as $(\ln N)^{-1}$. To determine these frequencies it is convenient to exploit the separability of the integrable dynamics by introducing the Jacobi elliptic coordinates $u_\nu$, $\nu=0,\dots,\mathscr{N}$, from which the action-angle variables and the corresponding frequencies can be computed exactly~\cite{jacobi1884cgj}.} These are defined as the zeros of the function
\begin{equation} \label{eq:ul}
    \varphi(u) = \ 1 - \frac{\lambda}{2N} \sum_{\nu=0}^{\mathscr{N}} \frac{|\bar{\eta}_\nu|^2}{u-\omega_\nu^2} = \prod_{\nu=0}^\mathscr{N} \frac{u-u_\nu}{u-\omega_\nu^2} \, , 
\end{equation}
while their relative momenta are defined as
\begin{equation}
 P_\nu = \frac{1}{2 \lambda} \partial_t \varphi(u)|_{u = u_\nu}  =  \frac{1}{2\lambda}  \frac{ \dot{u}_\nu}{u_\nu-\omega_{\nu}^2}  \prod_{\mu \neq \nu} \frac{u_\nu - u_{\mu}}{u_\nu - \omega_{\mu}^2} \,
\end{equation}
(see App.\,\ref{app:sperability} for the details). Due to the form of the function $\varphi(u)$, $u_\nu$ is such that $\omega_\nu < u_\nu < \omega_{\nu+1}$ for all $\nu \neq \mathscr{N} - 1$, while $u_\mathscr{N}$ is confined in the semi-axis $u > \omega_\mathscr{N} = 1$. In particular, in the case $\mathscr{N} = 1$ we recover the classical degree of freedom as $u_1 (t) = 1 + \lambda \xi^2(t)/2$ . As shown in App.~\ref{app:sperability}, see also Refs.~\cite{wojciechowski1985integrability,giachetti2025universality}, the set $u_\nu$, $P_\nu$ is indeed separable, as
\begin{equation} \label{eq:Pu}
\lambda^2 P_\nu^2 + V(u_\nu) = 0    
\end{equation}
where $V(u)$ is the potential
\begin{equation} \label{eq:V(u)}
    V(u) = u + r -  \frac{\lambda}{N} \sum_{\nu=0}^{\mathscr{N}} \frac{\epsilon_\nu}{u-\omega^2_\nu} + \frac{\lambda^2}{4N}\sum_{\nu=0}^{\mathscr{N}} \frac{\ell_\nu^2}{(u-\omega^2_\nu)^2}  
\end{equation}
($\ell_{\nu =0}$ is implied to be identically zero). Since the motion takes place within the regions $V(u) < 0$ ($u>0$) the inversion points $u_\nu^{\pm}$ of each $u_\nu$ coincide with the positive zeros of $V(u)$. Thus, the energy of the quasi-particles $u_\nu$ oscillates between the inversion points $u_\nu^{\pm}$ (see Fig.\,\ref{fig:Potential}).  

\subsection{Spectrum of the resonances}

\begin{figure}
    \centering
    \includegraphics[width=0.85\textwidth]{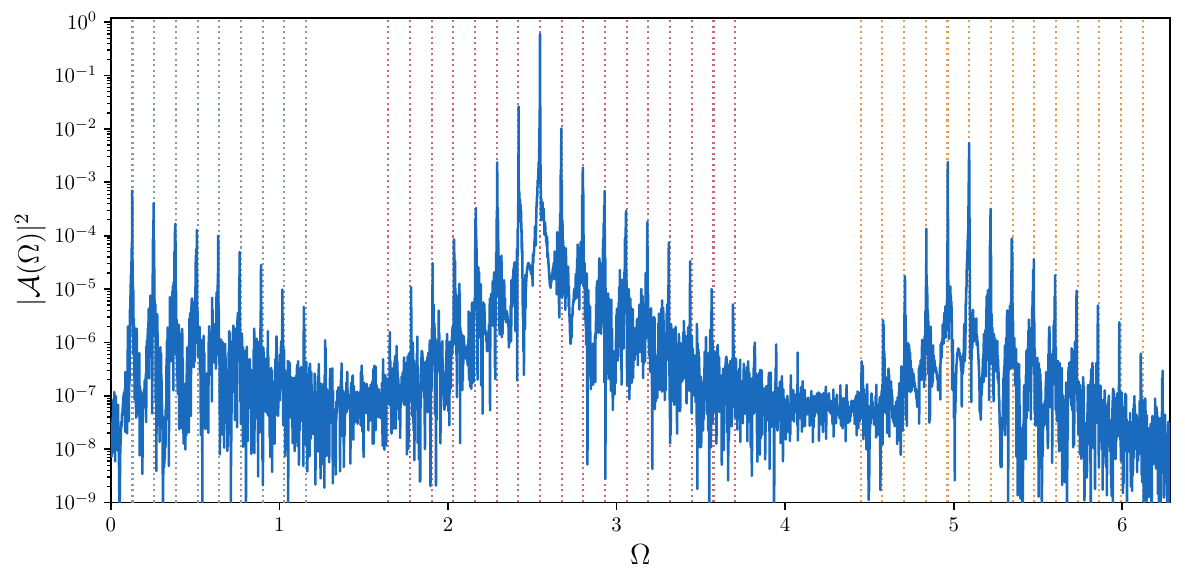}
    \caption{Spectral density of $\mathcal{A}(t) = m^2(t) - \overline{m^2}$ ($\overline{\cdot}$ representing the time average), for the single-resonance ($\mathscr{N}=1$) phase. The location of the peaks is compared with the expected quasi-periodic comb-structure $p_0 \ \Omega_0 + p_1 \ \Omega_1$, $p_0, p_1 \in \mathbb{Z}$ (dashed vertical lines), finding an excellent agreement. Here $\Omega_{1}$, corresponding to the most prominent peak, coincides with the frequency of the classical motion in Eq.\,\eqref{eq:quartic_potential}, while $\Omega_0 \sim (\ln N)^{-1}$ is the resonant frequency in Eq.\,\eqref{eq:Omeganu}. Peaks close to the different harmonics $p_1 \Omega_1$ of the classical frequency are highlighted in different colors. \textcolor{black}{The spectrum is obtained by numerically integrating the reduced resonant Hamiltonian rather than the full lattice dynamics. The discrete Fourier transform $\mathcal{A}(\Omega)$ is computed over the time window $0\leq t\leq t_{\rm max} = 5000$, sampled with time step $\Delta t = 0.05$. The plotted spectral density is the normalized power spectrum $|\mathcal{A}(\Omega)|^2/T^2$, where $T=t_{\rm max}/\Delta t+1$ denotes the total number of sampled time points.} Here we considered a quench from $m_{\rm gs} = 6$ to $r=-2$ with $\lambda = 1$, $N = 10^{14}$. Due to the logarithmic scaling in $\Omega_0$, such a large value of $N$ is needed both to have a clear scale separation and to get rid of finite-size effects. \textcolor{black}{Consequently, the dynamics is obtained by numerically integrating the equations of motion of the reduced Hamiltonian \eqref{eq:mathcalHRes}.}} \label{fig:SpectrumRes}
\end{figure}
As the system is integrable also beyond the linear regime, the dynamics of the resonant modes can be written as the superposition of $\mathscr{N}$ frequencies $\lbrace \Omega_\nu \rbrace$. Of these $\mathscr{N}$ are of order $(\ln N)^{-1}$, as they correspond to the resonances, while the last one will correspond to the classical mean-field frequency. 

However, these frequencies do not correspond to the ones of a particle in the potential $V(u)$ as $P_\nu \neq \dot{u}_\nu$, and the $u_\nu$ will not be periodic. Instead, the frequencies can be computed by expressing $\mathcal{H}_\mathscr{N}$ in terms of the action variables 
\begin{equation} \label{eq:Jnu}
    J_\nu = \oint \frac{du_\nu}{2 \pi} P_\nu =  \int^{u_\nu^+}_{u_\nu^-} \frac{du}{\pi \lambda} \sqrt{-V(u)} .
\end{equation}
and by computing $\Omega_{\nu} = \partial \mathcal{H}/\partial J_\nu$. 

From Eq.~\eqref{eq:V(u)} it follows that, close to the poles of $V(u)$ one has 
\begin{equation}
    V(u) = \frac{\lambda^2 \ell_\nu^2}{4N^2(u-\omega_\nu^2)^2} - \frac{\lambda \epsilon_\nu}{N(u-\omega_\nu^2)} + \Delta_\nu + O(u-\omega_\nu)
\end{equation}
where $\Delta_\nu$ is the same stability parameter defined in Eq.\,\eqref{eq:Deltanu}. As shown in Fig.\,\ref{fig:Potential}, the resonant modes with $\Delta_\nu < 0$ correspond to finite amplitude oscillations, while off-resonant modes result in oscillations with vanishing amplitude, whose effect is negligible at large $N$. Instead, the inversion points $u_\nu^\pm$ of the resonant modes $\nu = 0, \cdots, \mathscr{N}-1$ will approach $\omega_\nu^2$ and $\omega_{\nu+1}^2$ respectively in the thermodynamic limit. As shown in App.\,\ref{app:action}, at the leading order in $N$ one has 
\begin{equation}
    \mathcal{H}_\mathscr{N} = \bar{H}(J_\mathscr{N}) + \frac{2 \pi}{\ln N} \sum_{\nu=0}^{\mathscr{N}-1} J_\nu \sum_{\mu = 0}^{\nu} \sqrt{|\Delta_\mu|} 
\end{equation}
where $\bar{H}(J_\mathscr{N})$ corresponds to the Hamiltonian of the classical degree of freedom. Finally, for the resonant modes $\nu = 0, \cdots, \mathscr{N}-1$, one can write
\begin{equation}   \label{eq:Omeganu}
    \Omega_\nu = \frac{2 \pi}{\ln N} \sum_{\mu = 0}^{\nu} \sqrt{|\Delta_\mu|} \, .
\end{equation}
As expected, we find a quasi-periodic motion in which the frequency $\Omega_\mathscr{N} = O(1)$ of the classical oscillations of $\bar{\eta}$, coexists with the resonant frequencies $\Omega_\nu \sim (\ln N)^{-1}$, $\nu = 0, \dots, \mathscr{N}-1$. This prediction has been compared with the numerics for the single resonance case $\mathscr{N}=1$ in Fig.~\ref{fig:SpectrumRes} where the spectral density of the observable $m^2(t)$ is analyzed. 
This case corresponds to an unbroken effective permutational symmetry. Thus, a clear scale separation is observed, as the dynamics is the superposition of two different frequencies and the classical motion can be seen as a fast beat on top of the slow resonant dynamics. The resonant mode $|\bar{\eta}_0|$ will periodically return small, with a period $\Omega_0^{-1} \sim \ln N$. 

\section{Correlations and entanglement spreading}  \label{sec:entanglement}
\textcolor{black}{Having characterized the dynamics of the resonant modes, we now turn to its physical consequences. In particular, we investigate how resonances generate spatial quantum correlations and entanglement, gradually driving the system away from the classical phase. In the following we focus on the large-$N$ regime, and we assume that the resonant modes are active, i.e. they have reached macroscopic occupations $\bar{\eta}_\nu = O(1)$.} As the mean-field degree of freedom $\bar{\eta}$ is linked to one-site dynamics, we expect quantum correlation between different sites only to arise due to resonances. This expectation is indeed confirmed by the analysis of Ref.~\cite{giachetti2021entanglement}. In this section, we revise the link between entanglement spreading and the presence of resonances and derive semi-analytical expressions for the entanglement spreading. The presence of resonances thus gradually drives the system away from the mean field description, which is completely washed away in the limit of an extensive $\mathscr{N} \sim N$. 

As a measure of the quantum spatial correlations we will consider the entanglement entropy $S$ associated with a bipartition of the system into an interval $A$ of length $L_A$ and its complement. It is defined as $S = - \tr \varrho_A \ln \varrho_A$, where $\varrho_A = \tr_{\bar{A}} \varrho$ is the reduced density matrix associated with the subsystem $A$. In the case of periodic boundary conditions, the exact position of the interval is not relevant. The entropy $S(t)$ can be in turn expressed in terms of the two-point correlations of the bosonic operators $\Phi_j$, $\Pi_j$,
\begin{equation} \label{eq:correlation}
\gamma = \text{Re} \begin{pmatrix} \me{\Phi_j (t) \Phi_{j'}(t)} & \me{\Phi_j (t) \Pi_{j'}(t)} \\  
\me{\Pi_j (t) \Phi_{j'}(t)} & \me{\Pi_j (t) \Pi_{j'}(t)}
\end{pmatrix} 
\end{equation}
by following the procedure of Ref.~\cite{casini2009entanglement}. This is a consequence of two facts: $(i)$ for a ground state quench the initial state $\varrho(0)$ is a bosonic Gaussian state, $(ii)$ in the $n \rightarrow \infty$ limit, the state of system remains Gaussian at any time $t$. In particular, we have $S(t) = \sum_n s(\sigma_n)$ where 
\begin{equation} \label{eq:entropyspectrum}
s(\sigma) = \left( \sigma + \frac{1}{2} \right) \ln \left( \sigma + \frac{1}{2} \right) - \left( \sigma - \frac{1}{2} \right) \ln \left( \sigma - \frac{1}{2} \right) ,  
\end{equation}
and $\sigma_n$ are the symplectic eigenvalues of the matrix $\gamma$ in Eq.~\eqref{eq:correlation}, reduced on the subsystem $j,j' \in A$. More explicitly, if we denote such a reduced correlation matrix as $\gamma_{\rm red}$, the $\sigma_n$ are a set of $L_A$ positive numbers which are eigenvalues of  $ \textit{i}  \mathbb{J} \gamma_{\rm red} $, $\mathbb{J}$ being the symplectic identity
\begin{equation}
\mathbb{J} = \begin{pmatrix}
0 & \mathbb{I}_{L_A} \\ -\mathbb{I}_{L_A} & 0
\end{pmatrix} \, .
\end{equation}
In the limit of large size $N$, the matrix $\gamma_{\rm red}$ will be dominated by the classical mode $\bar{\eta}$ and the resonant ones, as the latter becomes $O(\sqrt{N})$ on $t \sim \ln N$. More explicitly 
\begin{equation} \label{eq:CorrelationRes}
\begin{split}
\me{\Phi_j \Phi_{j'}}  = |\bar{\eta}^2| \ \delta_{j'j} +  \sum_{\nu=0}^{\mathscr{N}-1} &( |\bar{\eta}^2_\nu|  - N^{-1} |\bar{\eta}^2| ) e^{2 \pi \text{i} \nu (j-j')/N}, \hspace{0.2cm}
\me{\Pi_j  \Pi_{j'}}= |\bar{p}^2| \ \delta_{j'j} + \sum_{\nu=0}^{\mathscr{N}-1} ( |\bar{p}^2_\nu|  - N^{-1} |\bar{p}^2| ) e^{2 \pi \text{i} \nu (j-j')/N} , \\
\Re \me{\Pi_j  \Phi_{j'}} &= \Re(\bar{p}^* \bar{\eta}) \  \delta_{j'j} + \sum_{\nu=0}^{\mathscr{N}-1} (\Re(\bar{p}_\nu^* \bar{\eta}_\nu) - N^{-1} \Re(\bar{p}^* \bar{\eta}) ) e^{2 \pi \text{i} \nu (j-j\prime)/N} .
\end{split}    
\end{equation}
Different regimes appear
\begin{itemize}
    \item \textit{Case} $\mathscr{N} = 0$: in this case the correlations in Eqs.~\eqref{eq:CorrelationRes} are all diagonal and $ \sigma^2_n = |\bar{\eta}|^2 |\bar{p}|^2 - \Re(\bar{p}^* \bar{\eta})^2 = \Im (\bar{p}^* \bar{\eta}) = \ell^2$, so that has $\sigma_n = \ell$ for any $n=1, \cdots, L_A$.  As $S(t)$ remains constant, no entanglement is generated. Since for a ground state quench $\ell = 1/2$, $S(t) \equiv 0$ as expected from  Eq.~\eqref{eq:entropyspectrum}.  
    \item \textit{Case} $\mathscr{N} = 1$: in this case we have 
    $\me{\Phi_j \Phi_{j'}}  = |\bar{\eta}^2| \delta_{j'j} +  |\bar{\eta}^2_0| - N^{-1} |\bar{\eta}^2|$ (and analogously for the momentum and cross correlations). Only for $n=1$, $\sigma_n\neq 1/2$ and the entanglement contribution is given by 
    \begin{equation}
    \sigma_1  = \frac{1}{2} \sqrt{1 - 2 \rho (1-\rho) + 4 L_A (1-\rho) \chi(t) }
    \end{equation}
    where $\rho = L_A/N$ and 
    \begin{equation}
       \chi_1 (t) =  |\bar{p}_0^2 \bar{\eta}_1^2| + |\bar{p}^2_1 \bar{\eta}_{0}^2| - 2 \Re(\bar{p}_1^* \bar{\eta}_1) \Re(\bar{p}_{0}^* \bar{\eta}_{0}) 
   \end{equation}
    represents the overlap between the resonant mode $\bar{\eta}_0$ and the classical one $\bar{\eta}_1 \equiv \bar{\eta}$. \textcolor{black}{We find then
    \begin{equation}
        S = s(\sigma_1) \, .
    \end{equation}
    So far, the expression is exact:} let us notice that $S(t) \approx 0$ at any time in which the resonance returns inactive ($\eta_0  = O(1)$, $\chi_1 = O(N^{-1})$), so that quantum correlations follow a quasi-periodic pattern, on a timescale $\sim \ln N$. \textcolor{black}{We will now derive the leading term in $S(t)$ in the regime $L_A \gg 1$: in particular, we have to separate the case of an extensively long interval (i.e. $N \gg 1$ while keeping $\rho = O(1)$), and the case in which the interval $L_A \gg 1$ is kept finite as we increase $N$ (i.e. $1 \ll L_A \ll N$ and $\rho = O(N^{-1})$). In both cases,  $\sigma_1 = O(L_A^{1/2})$, provided that that the resonance is active, $\chi_1(t) = O(1)$. Thus, for an extensive interval we find}
    \begin{equation}
         S(t) \sim \frac{1}{2} \ln [L_A   (1- \rho) \chi_1 (t)] \, ,
    \end{equation}
    \textcolor{black}{while for the case $1 \ll L_A \ll N$ instead 
    \begin{equation}
         S(t) \sim \frac{1}{2} \ln [L_A   \chi_1 (t)] \, 
    \end{equation}
   (let us notice that both expression breaks down whenever the resonance is non active and $\eta_0 = O(1)$).}  The maximum entangled generated is thus of order $1/2 \ln L_A$, as observed also for long-range spin systems~\cite{lerose2020origin}.  
    \item \textit{Case} $\mathscr{N}  > 1$: here the permutational symmetry is effectively broken and the correlations acquire a spatial modulation, making an explicit computation of $S(t)$ impossible. However, close expressions can be derived in the regime $L_A \gg 1$. 
    Indeed, while the spatial modulation of the modes can no longer be ignored for $L_A = O(N)$ ($\rho = O(1)$), a principal component analysis of the $L_A$ by $L_A$ correlation matrix $\gamma_{\rm red}$ reveals that only a number $\lfloor \mathscr{N}/2 \rfloor$ of $O(L)$ eigenvalues $\sigma_n$ actually contributes to the entropy. In particular, as shown in App.\,\ref{app:entanglement},  \textcolor{black}{for an extensive interval} we have 
    \begin{equation}\label{eq:SGamma}
        S(t) \sim \frac{1}{2} \ln[ \text{pdet} \ \Gamma(t)]
    \end{equation}
    where $\text{pdet}$ denotes the pseudo-determinant of the $\mathscr{N}$ by $\mathscr{N}$, skew symmetric matrix
    \begin{equation} \label{eq:Gammat}
        \Gamma_{\mu  \nu} =  N \frac{\sin \left( \pi \rho (\mu-\nu)\right)}{\pi (\mu - \nu)} \left( \bar{\eta}_{\mu} \bar{p}_\nu - \bar{\eta}_{\nu} \bar{p}_\mu \right) \, .
    \end{equation}
    For $\mathscr{N} = 2$ one has 
    \begin{equation}
        S(t) \sim \ln[ L_A  \left( \bar{\eta}_{1} \bar{p}_2 - \bar{\eta}_{2} \bar{p}_1 \right) \text{sinc}(\pi \rho) ]
    \end{equation}
    while for $\mathscr{N} = 3$
    \begin{equation}
        S(t) \sim \frac{1}{2} \ln \left[  L^2_A \sum^{2}_{\mu > \nu}  (\bar{\eta}_{\mu} \bar{p}_\nu - \bar{\eta}_{\nu} \bar{p}_\mu)^2  \text{sinc}^2 (\pi (\mu - \nu)\rho) \right] \, . 
    \end{equation}
    \textcolor{black}{Finally, in the regime $1 \ll L_A \ll N$ (i.e. $\rho = O(N^{-1})$)} one can derive the expression
    \begin{equation}
        S(t) \sim \frac{1}{2} \ln  \left[L^2_A \sum^{\mathscr{N}-1}_{\mu > \nu}  (\bar{\eta}_{\mu} \bar{p}_\nu - \bar{\eta}_{\nu} \bar{p}_\mu)^2 \right]
    \end{equation}
    valid for all $\mathscr{N} > 1$. Let us notice that in this case the maximum amount of entanglement generated is of order $S(t) \sim \ln L_A$: this can be traced back to the fact that for $\mathscr{N} >1$ the leading term is given by the overlap of two different resonant modes, while for $\mathscr{N} =1$ by the overlap of the single resonant mode $\bar{\eta}_{0}$ and the classical mode $\bar{\eta}$. \textcolor{black}{These expressions we derived are valid whenever at least two of the resonant modes are active; on the other hand, as the motion is quasi-periodic, for $\mathscr{N} > 2$ this will happen for every $t > t_{\rm Ehr} \sim \ln N$:} in this case, as the dynamics of the resonant mode is given by the interference of different frequencies, $S(t)$ oscillates around a finite value, as a finite amount of entanglement is generated.
\end{itemize}

In summary, quantum resonances are the mechanism that spreads correlations in strong long-range systems, disrupting the mean-field dynamics. Since $S \sim c \ln L_A$, entanglement never obeys  volume-law as in the short-range limit, in line with the general understanding of strong long-range dynamics~\cite{pappalardi2018scrambling,lerose2020origin,giachetti2021entanglement}. In addition, our analysis reveals the importance of the multi-resonant regime: the presence of $\mathscr{N} \geq 2$ spatially-modulated active resonances results in an enhanced entanglement production ($c= 1$ versus $c=1/2$ of the single active resonance case). Multi-resonances are also associated with a stable production of entanglement due to the interference among different resonant modes. 

\section{Conclusions}

\textcolor{black}{In this work, we developed a quantitative theory of the mesoscopic dynamics of the strong long-range quantum $O(n)$ model in the large-$n$ limit. Building on the phenomenological picture of Ref.~\cite{giachetti2021entanglement} and on the integrable formalism introduced in Ref.~\cite{giachetti2025universality}, we showed how the interplay between a classical mean-field degree of freedom and a finite set of quantum modes gives rise to a nontrivial finite-size dynamics on the logarithmic timescale $t_{\rm Ehr}\sim \ln N$. The key mechanism behind this behavior is the activation of a finite number $\mathcal N$ of parametric resonances, induced by the discrete structure of the long-range spectrum.}

More specifically, the integrals of motion allow us to characterize the onset of resonances in the linear regime, construct the associated phase diagram, and derive an effective Hamiltonian for the nonlinear dynamics of the resonant sector. In turn, this reduced description gives access to the quasiperiodic spectrum of the mesoscopic dynamics and provides a unified interpretation of finite-size effects in strong long-range systems.

From this perspective, the departure from the classical phase can be understood as a progressive proliferation of resonant modes. While in the strict thermodynamic limit the dynamics is effectively described by an ultra-local classical degree of freedom and retains an emergent permutational invariance, multiple resonances destroy this picture on the timescale $t_{\rm Ehr}$. The same resonant sector is also responsible for the build-up of quantum correlations and entanglement. Although the quasi-classical  nature of the dynamics prevents the onset of volume-law entanglement, we find that the presence of several resonant modes enhances both the rate and the stability of entanglement production.

More broadly, our results highlight the large-$n$ $O(n)$ model as a particularly transparent setting in which mesoscopic long-range dynamics can be analyzed beyond a semi-qualitative Floquet description. In this sense, the exact solution of the model provides a useful benchmark for future analytical approximations—such as the $1/n$ expansion~\cite{ZinnJustin}—and for the study of more generic long-range systems where the same interplay between persistent collective oscillations and spatially modulated quantum modes is expected to arise.

A natural extension of our analysis concerns the crossover from the strong to the weak long-range regime, $\alpha\to d$, where the resonance mechanism and the associated separation of timescales are expected to break down. In this respect, it would be particularly interesting to clarify the relation between the ultra-local oscillations discussed here and the persistent high-energy oscillations observed in the short-range $O(n)$ model~\cite{maraga2015aging}, which have recently been related to Higgs-like modes lying beyond the edge of the spectral band~\cite{giachetti2025universality}.

Finally, the present framework could be expanded to explain the interplay between persistent mean-field oscillations and spatially-modulated quantum modes observed in other strong-long-range models—most notably in semiclassical analyses of the LMG model~\cite{lipkin1965validity} (see e.g.  Refs.~\cite{lerose2018chaotic,pappalardi2018scrambling}). As the interplay between semiclassical chaos and emergent integrability in long-range many-body systems remains largely unexplored, clarifying whether such oscillations can be universally interpreted as manifestations of resonant or near-integrable sectors would significantly unify the phenomenology of long-range models. 

\section{Acknowledgment} 
GG acknowledges the support of the MSCA Grant 101152898 (DREAMS). This research was funded by the Swiss National Science Foundation (SNSF) grant numbers 200021--207537 and 200021--236722, by the Deutsche Forschungsgemeinschaft (DFG, German Research Foundation) under Germany's Excellence Strategy EXC2181/1-390900948 (the Heidelberg STRUCTURES Excellence Cluster) and by the European Union under GA No. 101077500–QLRNet. Partial support by grant NSF PHY-230935 to the Kavli Institute for
Theoretical Physics (KITP) is also acknowledged.

\appendix

\section{Floquet theory for the Floquet-Hill equation}
\label{app:Floquet}
We remind briefly the basics of the Floquet theory with special attention to the case of the Floquet-Hill equation
\begin{equation}
    \ddot{\eta} (t) + a_\nu (t) \ \eta(t) = 0 ,
\end{equation}
with $a(t)= a(t+\tau)$, which corresponds to the equation of motion of each $\eta_\nu$ \eqref{eq:FloquetHill} within the linear regime. 

We may now consider the set of two (independent) fundamental solutions  $\eta_1(t)$, $\eta_2(t)$ such that $\eta_1(0) = 1$, $\dot{\eta}_1 (0) = 0$ and $\eta_2(0) = 0$, $\dot{\eta}_2 (0) = 1$ respectively. The Wronskian of the solutions $W = \eta_1 \dot{\eta}_2 - \eta_2 \dot{\eta}_1$ is conserved. Moreover, being $a(t)$ periodic of period $\tau$, $\eta_1(t+\tau)$, $\eta_2(t+\tau)$ for a new pair of independent solutions, and can thus be expressed as a linear combination of $\eta_1 (t)$, $\eta_2(t)$, namely
\begin{equation} \label{Flomatrix}
    \begin{pmatrix}
    \eta_1 (t+\tau) \\ \eta_2 (t+\tau)
    \end{pmatrix} = C \begin{pmatrix}
    \eta_1 (t) \\ \eta_2 (t)
    \end{pmatrix},
\end{equation}
where $C$ is a constant square matrix of order $2$, known as monodromy matrix. A constraint of the matrix $C$ comes by the observation that 
\begin{equation}
    W (t + \tau) = \det C \ W(t)
\end{equation}
so that, $\det C = 1$ as $W(t)$ is constant. 

\textcolor{black}{It is convenient to diagonalize the matrix $C$: we introduce thus two independent linear combinations $\eta_{\pm} (t)$, of $\eta_{1,2} (t)$ such that $\eta_{\pm} (t+\tau) = \Lambda_{\pm} \eta_{\pm} (t)$, $\Lambda_{\pm}$ being the eigenvalues of $C$. On the other hand, since $C$ is real and $\det C = 1$ we have that only two cases are possible, either $\Lambda_{\pm} = e^{\pm \text{i} \varpi \tau}$ ($|\tr C| > 2$) or $\Lambda_{\pm} = e^{\pm \lambda \tau}$ ($|\tr C| < 2$) for some real (positive) $\lambda$, $\varpi$ (we are here neglecting the boundary case $|\tr C| =2$). In this two cases we can thus write $\eta_{\pm} (t) = f_{\pm} (t) e^{\pm \text{i} \varpi t }$, or $\eta_{\pm} (t) = f_{\pm} (t) e^{\pm \lambda t }$ respectively, with $f_{\pm} (t + \tau) = f_\pm (t)$. }

\textcolor{black}{As the generic solution of the Floquet-Hill equation can be written as a linear combination of $\eta_{1,2}$, and thus of $\eta_\pm$ one has that each solution take the form 
\begin{equation} \label{eq:FHoffres}
    \eta(t) = f_+ (t) e^{\text{i} \varpi t} + f_- (t) e^{-\text{i} \varpi t} , 
\end{equation}
or 
\begin{equation} \label{eq:FHres}
    \eta(t) = f_+ (t) e^{\lambda t} + f_- (t) e^{-\lambda t}  . 
\end{equation}
While in the former case we have a quasi-periodic solution with $\eta(t) \sim O(1) \ \forall t$, in the latter we have a parametric resonance and $\eta \sim e^{\lambda t}$.}

\textcolor{black}{Notice how the expression \eqref{eq:etalinear} for the $\eta_\nu$ derived in the linear regime by exploiting the integrability condition naturally falls into the class of Eq.\,\eqref{eq:FHoffres} or Eq.\,\eqref{eq:FHres} depending on the sign of $\Delta_\nu$.  }

\textcolor{black}{If the time coordinate is replaced by a spatial one, the Floquet-Hill equation becomes formally equivalent to the Schrodinger equation in a periodic potential. In this case the resonances corresponding now to the forbidden solution, i.e. the gap between different bands; while the quasi-periodic ones are Bloch waves. }

\section{Separability of the Jacobi coordinates}
\label{app:sperability}
Let us set $\eta_{\nu} = \xi_\nu e^{i \theta_\nu}$: as $\theta_\nu$ is cyclic, one has that $\theta_\nu$, $\ell_\nu$ are trivially a pair of separable canonical coordinates. The Jacobi coordinates can be thought as a point canonical transformation from the $\xi_\nu$ to the $u_\nu$, defined as
\begin{equation} 
    \varphi(u) = \ 1 - \frac{\lambda}{2N} \sum_{\nu} \frac{\xi_\nu^2}{u-\omega_\nu^2}  = \prod_{\nu} \frac{u-u_\nu}{u-\omega_\nu} \, .
\end{equation}
We will now prove that the $u_\nu$ are a set of separable coordinates for $\mathcal{H}_\mathscr{N}$ in \eqref{eq:mathcalHRes}. 

First we notice that, by replacing the equations of motion in $\partial_t^2 \phi$ and by making use of the elementary identity 
\begin{equation}
    \frac{1}{(u-u_\nu)(u-u_\mu)} = \frac{1}{u_\nu-u_\mu} \left( \frac{1}{u-u_\nu} - \frac{1}{u-u_\mu} \right)
\end{equation}
(with $\nu \neq \mu$) we find 
\begin{equation}
    V(u) = (u+m^2) \varphi^2 + \frac{1}{2} \varphi \ \partial^2_t \varphi(u) - \frac{1}{4} (\partial_t \varphi)^2 \ ,
\end{equation}
which for $u = u_\nu$ becomes
\begin{equation}
     V(u_\nu) + \frac{1}{4} (\partial_t \varphi)^2|_{u=u_\nu} = 0 \, .
\end{equation}
To prove Eq.\,\eqref{eq:Pu}, and thus the separability of the set $\lbrace u_\nu \rbrace$ one has to show that the momentum conjugate to $u_\nu$ is $P_\nu =(2 \lambda)^{-1} \ \partial_t \varphi|_{u=u_\nu}$. 

The momentum $P_\nu$ conjugate to $u_\nu$ is defined as  
\begin{equation}
    P_\nu = \sum_\mu \frac{\partial \xi_\mu}{\partial u_{\nu}} p_\mu 
\end{equation}
It is thus convenient to write explicitly the $\xi_\nu$ in terms of the $u_\nu$, which can be done by as the $\xi_\nu$ are linked to the residues of the poles of $\varphi(u)$. We get
\begin{equation}
    \frac{\lambda}{4} \xi_\nu^2 = (\omega_\nu^2 - u_\nu) \prod_{\omega_\nu \neq \omega_\mu} \frac{\omega_{\nu}^2 - u_\mu}{\omega_{\nu}^2-\omega_\mu^2} \, , 
\end{equation}
from which
\begin{equation}
    \frac{1}{\xi_\nu} \frac{\partial \xi_{\nu}}{\partial u_\mu} = \frac{1}{2} \frac{1}{\omega_\nu^2-u_\mu} \, , 
\end{equation}
 and finally
\begin{equation}
    P_\nu = \sum_\mu \frac{\partial \xi_\mu}{\partial u_{\nu}} p_\mu = \frac{1}{2} \sum_\mu \frac{\xi_\mu p_\mu}{\omega_\mu^2 - u_\nu} = - \frac{1}{2 \lambda} \partial_t \varphi|_{u = u_\nu} 
\end{equation}
where we used the equation of motion $\dot{\xi}_\nu = p_\nu$ for the original variables. This proves the separability.

Let us notice that, more explicitly, one has
\begin{equation} \label{eq:p(qdot)}
    P_\nu =  \frac{N}{2\lambda}  \frac{ \dot{u}_\nu}{u_\nu-\omega_{\mu}^2}  \prod_{\nu \neq \mu} \frac{u_\nu - u_{_\mu}}{u_\nu - \omega_{_\mu}^2} \, . 
\end{equation}

\section{Computation of the action variables}
\label{app:action}
We want now to compute the action variables $J_\nu$ in Eq.\,\eqref{eq:Jnu}. Let us first notice that, far from the poles $u = \omega^2_\nu$ of the potential $V(u)$, in the limit $N \rightarrow \infty$ we have 
\begin{equation}
    V(u) \approx \tilde{V} (u) = u + r - \frac{\lambda \epsilon_\mathscr{N}}{u-1} + \frac{\lambda^2 \ell^2_\mathscr{N}}{4(u-1)^2}
\end{equation}
up to $O(N^{-1})$, while close to the poles
\begin{equation}
    V(u) \approx \Delta_\nu - \frac{\lambda \epsilon_\nu}{N(u-\omega_\nu^2)} + \frac{\lambda^2 \ell^2_\nu}{4N^2(u-\omega_\nu^2)^2} \, .
\end{equation}
On the other hand, in this limit $u_{\nu}^{-} = \omega_\nu^2 + O(N^{-1})$, $u_{\nu}^{+} = \omega_{\nu+1}^2 + O(N^{-1})$ for $\nu = 0, \dots,\mathscr{N}-1$, so that we have
\begin{equation}
    J_\nu = \int^{\omega_{\nu+1}^2}_{\omega_{\nu}^2} \frac{du}{\pi \lambda} \sqrt{-\tilde{V}(u)} + \frac{\ln N}{2 \pi N} \left( \frac{\epsilon_{\nu}}{\sqrt{|\Delta_\nu}|} - \frac{\epsilon_{\nu+1}}{\sqrt{|\Delta_{\nu+1}}|} \right) + O(N^{-1})
\end{equation}
for $\nu = 0, \cdots,\mathscr{N}-2$ while
\begin{equation}
    J_{\mathscr{N}-1} = \int^{u_{\mathscr{N}-1}^{+}}_{\omega_{\nu}^2} \frac{du}{\pi \lambda} \sqrt{-\tilde{V}(u)} + \frac{\ln N}{2 \pi N} \frac{\epsilon_{\mathscr{N}-1}}{\sqrt{|\Delta_{\mathscr{N}-1}|}} + O(N^{-1})
\end{equation}
and
\begin{equation}
    J_{\mathscr{N}} = \int^{u_{\mathscr{N}}^{+}}_{u_{\mathscr{N}}^{-}} \frac{du}{\pi \lambda} \sqrt{-\tilde{V}(u)} + O(N^{-1}) \, .
\end{equation}
We have that 
\begin{equation}
    \sum_{\mu = \nu}^{\mathscr{N}-1} J_\mu = \int^{u_{\mathscr{N}-1}^{+}}_{\omega_\nu^2} \frac{du}{\pi \lambda} \sqrt{-\tilde{V}(u)}  \ + \frac{\ln N}{2 \pi N} \frac{\epsilon_\nu}{\sqrt{|\Delta_\nu|}} + O(N^{-1}) 
\end{equation}
Let us notice that $\tilde{V}(u)$ depends only on $\epsilon_\mathscr{N}$: we have then that 
\begin{equation}
    J_\mathscr{N} = f_\mathscr{N} (\epsilon_\mathscr{N})
\end{equation}
while for $\nu = 0, \dots, \mathscr{N}-1$
\begin{equation}
    \frac{\epsilon_\nu}{N} = \frac{2 \pi \sqrt{|\Delta_\nu|}}{\ln N} \left(- f_\nu (\epsilon_\mathscr{N}) + \sum_{\mu = \nu}^{\mathscr{N}-1} J_\mu + O(N^{-1}) \right)  \, 
\end{equation}
(where $f_\nu$ are a set of function implicitly defined by the above expressions). Finally one has 
\begin{equation}
    \mathcal{H}_\mathscr{N} = \epsilon_{\mathscr{N}} + \frac{1}{N} \sum_{\mu=0}^{N-1} \epsilon_\mu = \bar{H}(J_\mathscr{N}) + \frac{2 \pi}{\ln N} \sum_{\nu=0}^{\mathscr{N}-1} J_\nu \sum_{\mu = 0}^{\nu} \sqrt{|\Delta_\mu|} + o(N^{-1})
\end{equation}
where at the leading order in $N$, $\bar{H}$ is the Hamiltonian of the collective mode $\bar{\eta}$.

\section{Entanglement entropy for the multi-resonant case}
\label{app:entanglement}
We want now to prove Eq.\,\eqref{eq:SGamma} for $S(t)$ in the multi-resonant case. In this case, the leading contribution to $S(t)$ comes from the overlap between resonances, so that we can neglect the contribution of the classical, mean-field mode $\bar{\eta}$. More formally, we define the
\begin{equation}
    \braket{j |\nu} = \frac{1}{\sqrt{N}} e^{2 \pi \text{i} \nu \ j/N} 
\end{equation}
with $j \in \lbrace 1,L \rbrace$, $\nu \in \lbrace0, \mathscr{N}-1 \rbrace$, corresponding to the spatial modulation of the resonances. While the vectors  are orthogonal on the whole system, on the subspace $A$ one has a finite overlap between resonances given by
\begin{equation}
     \braket{\mu |\nu} = \frac{\sin  (\pi \rho (\mu-\nu))}{\pi (\mu-\nu)} \, ,
\end{equation}
where once again $\rho = L_A/N$. Finally as $\Im (\bar{p}_\nu^* \bar{\eta}_\nu) = O(N^{-1})$, we choose $\bar{\eta}_\nu$, $\bar{\eta}_\mu$ to be real. One has thus 
\begin{equation}
    \gamma_{\rm red} =  N \begin{pmatrix}
    Q & R \\ R & P
    \end{pmatrix},
\end{equation}
with 
\begin{equation}
Q =   \sum_{\nu=0}^{\mathscr{N}-1} \bar{\eta}_\nu^2 \ket{\nu} \bra{\nu}, \hspace{.5cm}
P  =  \sum_{\nu=0}^{\mathscr{N}-1} \bar{p}_\nu^2  \ket{\nu} \bra{\nu}, \hspace{.5cm}
R = \sum_{\nu=0}^{\mathscr{N}-1} \bar{p}_\nu \bar{\eta}_\nu \ket{\nu} \bra{\nu} , 
\end{equation}
so that
\begin{equation}
    \text{i} \mathbb{J} \gamma_{\rm red} = \text{i} N \sum_\nu (\bar{p}_\nu \ket{\nu}_1+\bar{\eta}_\nu \ket{\nu}_2) (-\bar{\eta}_\nu \bra{\nu}_1+\bar{p}_\nu \bra{\nu}_2) \equiv \text{i} N \sum_\nu \ket{u_\nu} \bra{v_\nu}
\end{equation}
where $\ket{\nu}_{1,2}$ represents the vector basis acting on the two subspaces and $\ket{u_\nu} = \bar{p}_\nu \ket{\nu}_1 + \bar{\eta}_\nu \ket{\nu}_2$, $\ket{v_\nu} = -\bar{\eta}_\nu \ket{\nu}_1 + \bar{p}_\nu \ket{\nu}_2$. As the rank of $\sum_\nu \ket{u_\nu} \bra{v_\nu}$ is at most $\mathscr{N}$, we will have several null eigenvalues: within our approximation, these must be interpreted as $O(N^{-1})$ eigenvalues, which in turn corresponds to the $O(1)$ $\sigma_j$, which give a subleading contribution to $S(t)$. We will just focus then on the $O(N)$ eigenvalues coming from the non-zero eigenvalues: these will coincide with the eigenvalues of the $\mathscr{N}$ by $\mathscr{N}$ imaginary skew-symmetric matrix
\begin{equation}
    \text{i} \ \Gamma_{\mu \nu} \equiv  N \braket{u_\mu |v_{\nu}} =  N (\bar{\eta}_\mu \bar{p}_\nu - \bar{\eta}_\nu \bar{p}_\mu) |\braket{\mu | \nu}|^2 \, ,
\end{equation}
which correspond to Eq.\,\eqref{eq:Gammat}. The spectrum of $\Gamma_{\mu \nu}$ is given by $\lfloor \mathscr{N}/2 \rfloor$ pairs  $\lbrace \text{i} \ \sigma_\nu, - \text{i} \ \sigma_\nu  \rbrace$ (with $\sigma_\nu > 0$, and $\sigma_\nu = O(N)$) plus an isolated zero eigenvalue for odd $\mathscr{N}$. We have thus that
\begin{equation}
    S(t) = \sum_\nu s(\sigma_\nu) \approx \sum_{\nu }  \ln \sigma_\nu =  \frac{1}{2} \ln \text{pdet} \Gamma (t)
\end{equation}
where the pseudo-determinant $\text{pdet}$ is the product of all the non-zero eigevalues.

\end{document}